\documentclass{ieeeaccess}

\usepackage{booktabs, colortbl, amsmath, amsfonts, graphicx, multirow, url, cite}
\usepackage{algorithmic}
\usepackage{graphicx}
\usepackage{textcomp}
\usepackage[caption=false,font=footnotesize]{subfig}
\usepackage{xcolor, soul}
\sethlcolor{yellow}

\def\BibTeX{{\rm B\kern-.05em{\sc i\kern-.025em b}\kern-.08em
    T\kern-.1667em\lower.7ex\hbox{E}\kern-.125emX}}

\def\rhoseg{{$\rho_{seg}$}}
\def\fn{{$f_n$}}

\def\wawenetfr{$\text{WAWEnet}_{1}^{\text{FR}}$}
\def\wawenettwonr{$\text{WAWEnet}_{1}^{\text{2}}$}
\def\wawenetmultiobj{$\text{WAWEnet}_{7}$}
\def\wawenetmultisubj{$\text{WAWEnet}_{11}$}
\def\wawenetobj{$\text{WAWEnet}_{\text{O}1}$}
\def\wawenetsubj{$\text{WAWEnet}_{\text{S}1}$}
\def\wawenetfoursubj{$\text{WAWEnet}_{\text{S}4}$}

\definecolor{gray}{rgb}{0.7, 0.7, 0.7}

\begin{document}
\history{Received 21 September 2023, accepted 28 October 2023, Date of publication 6 November 2023, date of current version 13 November 2023.}
\doi{10.1109/ACCESS.2023.3330640}

\title{Wideband Audio Waveform Evaluation Networks: Efficient, Accurate Estimation of Speech Qualities}

\author{Andrew~A.~Catellier\authorrefmark{1},~\IEEEmembership{Senior Member,~IEEE}
        and~Stephen~D.~Voran\authorrefmark{1},~\IEEEmembership{Senior Member,~IEEE}
\address[1]{Institute for Telecommunication Sciences, Boulder, CO 80305 USA (e-mail: svoran@ntia.gov)}}

\markboth
{Catellier \headeretal: Wideband Audio Waveform Evaluation Networks: Efficient, Accurate Estimation of Speech Qualities}
{Catellier \headeretal: Wideband Audio Waveform Evaluation Networks: Efficient, Accurate Estimation of Speech Qualities}

\corresp{Corresponding author: Andrew A. Catellier (e-mail: andrew@thisisreal.net).}

\begin{abstract}
Speech quality and speech intelligibility can vary dramatically across the wide range of currently available telecommunications systems, devices, and operating environments. This creates a strong demand for efficient real-time measurements of quality and intelligibility.
Wideband Audio Waveform Evaluation Networks (WAWEnets) are convolutional neural networks (CNNs) that operate directly on wideband audio waveforms in order to produce evaluations of those waveforms.
In the present work these evaluations give qualities of telecommunications speech (e.g., noisiness, intelligibility, overall speech quality).
WAWEnets are no-reference networks because they do not require ``reference'' (original or undistorted) versions of the waveforms they evaluate. Our initial 2020 WAWEnets publication introduces four WAWEnets, and each emulates the output of an established full-reference speech quality or intelligibility estimation algorithm.
We have updated the WAWEnet architecture to be more efficient and effective. Here we present a single WAWEnet that closely tracks seven different quality and intelligibility values with per-segment correlations in the range of 0.92 to 0.96.
We create a second network that additionally tracks four subjective speech quality dimensions. We offer a third network that focuses on just subjective quality scores and achieves a per-segment correlation of 0.97. The performance of our WAWEnet architecture compares favorably to models with orders-of-magnitude more parameters and computational complexity. This work has leveraged 334 hours of speech in 13 languages, more than two million full-reference target values, and more than 93,000 subjective mean opinion scores.
We also interpret the operation of WAWEnets and identify the key to their operation using the language of signal processing: ReLUs strategically move spectral information from non-DC components into the DC component. The DC values of 96 output signals define a vector in a 96-D latent space, and this vector is then mapped to a quality or intelligibility value for the input waveform.

\end{abstract}


\titlepgskip=-21pt

\maketitle

\section{Introduction}

\IEEEPARstart{W}{ired} and wireless telecommunications options continue to proliferate and evolve. Speech quality and speech intelligibility can vary dramatically between systems and devices, and further variation is driven by changes in acoustic environment, network loading, and radio conditions. Efficient real-time measurements of received speech quality and intelligibly are invaluable. For decades, researchers have been developing such measurement tools as telecommunications systems, devices, and use cases continue to evolve. This can and has caused entirely new classes of speech impairments and existing measurement tools may then fail to give meaningful results. This scenario motivates the development of new tools that do give meaningful results in the current environment.

Measurement tools fall into two major classifications---full-reference (FR) and no-reference (NR). FR tools require reference (transmitted) and impaired (received) speech and are practical in off-line applications. Some examples of FR tools developed over the years can be found in \cite{QuackenbushText,BarkSpectralDistortion,PSQM,MNBpartI,TOSQA2001,PESQref2,PEMO,WBPESQ,ESTOI,POLQA,SIIB,ViSQOL3,WARPQ}.
The most effective FR tools apply psychoacoustic transformations to reference and impaired speech and then compare these internal representations in ways that mimic key attributes of human judgement.

No-reference (NR) tools operate on impaired speech alone and are much more practical for real-time, in-service monitoring.
NR tools eliminate the need for time alignment and the issue of comparisons, but they require an embodied model for how speech should sound---independent of what speech was sent.  This is a significant challenge, but success allows for practical real-time, in-service monitoring of speech quality or intelligibility.  Thus these tools have also been named ``single-ended,'' ``in-service,'' ``non-intrusive,'' or ``output only.'' Some examples of early NR tools are given in \cite{RFK1994,P.563,ANIQUE+}. 

\vspace{-2mm}
\subsection{Existing Machine Learning Approaches}
\label{subsec:introML}
As machine learning (ML) has become more powerful and accessible, numerous research groups have sought to apply ML to the audio domain. In \cite{Hershey2017}, image classification networks use spectrograms as inputs in order to classify the content of arbitrary audio signals. The work in 
\cite{
Falk2006,
Soni2016,
Hakami2017,
Spille2018,
Pardo2018,
Fu2018,
Huber2018,
Salehi2018,
Mittag2019,
Ooster2019,
Santos2019,
Lo2019,
catellier2019wenets,
Gamper2019,
WAWEnets,
Pedersen2020,
Simou2020,
Choi2020,
Mittag2020Interspeech,
Mittag2020InterspeechB,
Jia2020,
Saishu2021,
Liu2021,
Nylen2021,
Serra2021,
Leng2021,
Zhang2021,
ReddyICASSP2021,
Zheng2021,
VoranQomex2021,
MittagQomex2021,
ReganoQomex2021,
Nessler2021IS,
Zhang2021IS,
Marcinek2021IS,
Mittag2021IS,
DongICASSP2022,
ReddyICASSP2022,
Ryandhimas2023,
Kumar2023} describes ML-based approaches to NR quality estimation.
Some of these NR tools produce estimates of subjective test scores that report speech or sound quality mean opinion score (MOS) \cite{
Falk2006,
Soni2016,
Hakami2017,
Salehi2018,
Mittag2019,
Ooster2019,
Santos2019,
Gamper2019,
Mittag2020Interspeech,
Liu2021,
Serra2021,
Leng2021,
ReddyICASSP2021,
ReganoQomex2021,
Nessler2021IS,
Kumar2023}, naturalness \cite{Lo2019, Choi2020, Mittag2020InterspeechB}, listening effort \cite{Huber2018}, noise intrusiveness \cite{Nessler2021IS}, and speech intelligibility \cite{Spille2018,Pedersen2020}. The non-intrusive speech quality assessment model called NISQA \cite{Mittag2021IS} uses log-mel-spectrograms to produce estimates of subjective speech quality as well as four constituent dimensions: noisiness, coloration, discontinuity, and loudness. DNSMOS P.835 \cite{ReddyICASSP2022} predicts subjective speech, background noise, and overall qualities for noise suppression algorithms.

Other NR tools produce estimates of objective values, including FR speech quality values \cite{Fu2018, catellier2019wenets, WAWEnets, Jia2020, Zhang2021, Zhang2021IS,DongICASSP2022, Ryandhimas2023,Kumar2023}, FR speech intelligibility values \cite{catellier2019wenets, WAWEnets, Jia2020, Zhang2021, Marcinek2021IS,DongICASSP2022,Ryandhimas2023,Kumar2023}, speech transmission index \cite{Pardo2018}, codec bit-rate \cite{Zheng2021}, and detection of specific impairments, artifacts, or noise types \cite{Simou2020, Saishu2021, Nylen2021, Marcinek2021IS}. Some of these tools perform a single task and others perform multiple tasks. The work in \cite{CooperICASSP2022} shows that large ML models trained using self-supervised learning to perform speech recognition tasks can also be successfully adapted to estimate subjective scores of speech naturalness.

The works cited here cover a wide variety of ML architectures. They address application areas that include room acoustics (noise and reverberation), speech enhancers, telecommunications systems, and hearing aids. Each addresses one or more of
narrowband (NB) (nominally 300~Hz--3.5~kHz),
wideband (WB) (nominally 50~Hz--7~kHz),
super-wideband (SWB) (nominally 50~Hz--16~kHz),
or fullband (FB) (nominally 20~Hz--20~kHz)
speech, except for \cite{Simou2020,Zheng2021}, which address music. Recent work shows that NR-tools can measure the speech quality of a system input in spite of the fact that such tools can only access the system output. \cite{VoranQomex2021}.

It is typical to consider agreement with subjective test results as the ultimate goal for NR tools. For ML-based tools this requires large datasets of subjective test results (most commonly speech quality MOS values) for training and validation. Since sufficient datasets are rare and expensive to generate through laboratory testing, so crowd-sourced tests are becoming common. Joint training or transfer learning can leverage objective FR quality values \cite{Mittag2019,Gamper2019,Nessler2021IS} or impairment categories
\cite{ReganoQomex2021} alongside MOS values to maximize the benefit of those MOS values. Semi-supervised learning \cite{Serra2021} is also an effective way to compensate for scarce MOS values. 

The operation of the absolute category rating (ACR) MOS subjective test parallels the operation of NR objective tools.  In ACR MOS tests listeners hear speech impaired with artifacts and give ratings without comparing the speech to a reference signal. Although great care is often taken to achieve absolute results, this can be a significant challenge; thus, ACR MOS results are often most realistically considered to be relative, not absolute. The use of the scale in any given subjective test can depend on the conditions included in that test and multiple other factors \cite{P8002,P1401}. For example, a given condition might be be rated 4.0 when it appears in a test with lower quality conditions. But that same condition might be rated 3.0 when it appears in a test with higher quality conditions. Per-subject corrections are explored in \cite{Nessler2021IS}. A method to learn per-test correction factors that is given in \cite{MittagQomex2021} may be viewed as the ML version of the linear algebra solution given in \cite{INLSA}.  

Key considerations when using ML to develop NR tools are the total amount of data available, how that data is used to both train and test a model, and the homogeneity or diversity of the training and testing data.
The number of talkers, languages, and sources of impairment are also potentially important factors, depending on the application.
Because the speech quality measurement community has not yet settled on standardized datasets, some published work is backed by extensive data and rigorous testing while other work could be described as an initial proof-of-concept for some specific architecture or application area, backed by a domain-specific dataset.
For this reason three separate datasets with three different application areas are considered in this work.

\subsection{The WAWEnet Machine Learning Approach}
\label{subsec:introNovelty}
The most common ML approach applies ML to a set of features that form a spectral description, often one that uses a perceptual frequency scale (e.g. mel, gammatone, or third-octave bands) and that may use the log of the spectral magnitudes since the log function can very roughly approximate the human perception of loudness. Alternately, ML has been applied to features such as mel-frequency cepstral coefficients, delta MFCCs, perceptual linear prediction coefficients, pitch, voice activity, frame energy, and features used in the legacy NR tools \cite{P.563,ANIQUE+}.

One distinction of WAWEnets is that they apply ML directly to speech waveforms instead of applying it to features extracted from speech waveforms. Extracting features from waveforms is a data compression step; it reduces the number of inputs per unit time the ML algorithm must process.  Good features identify and preserve all useful information in the speech waveform so that the ML algorithm may more easily and efficiently map that information to the desired target value. With WAWEnets there is no need to consider the pros and cons of various features---all information is made available to the network and the network itself effectively creates the features it needs. 
While there is no doubt that many different feature-based approaches have been successful in this problem space, WAWEnets demonstrate that using waveforms directly is a practical and effective alternative. 
We note that processing complex spectral values is another way to sidestep feature selection and preserve all of the speech signal information.  This has been explored very recently for speech assessment \cite{Ryandhimas2023} and previously studied in related fields \cite{WangICASSP2017,Trabelsi2018,Guimaraes2022}.

Before our initial WAWEnet publication \cite{WAWEnets} in 2020, we had found just one other NR research effort that considered speech waveforms for input to ML. In \cite{Pardo2018} ML is applied to a spectrogram representation of audio in order to estimate speech transmission index. But this spectrogram representation is also learned---waveforms are processed with 128 filters, each with length 128, and those filters are ``initialized with a Fourier basis (sine waves at different frequencies).'' So in effect, ML is actually applied to the audio waveform, but with a strong suggestion that the first ML step should be to calculate a spectral representation. This spectral representation does not use the magnitude function and thus it has the potential to preserve all of the information in the audio waveform.

We are aware of  some additional waveform-related research efforts that have emerged since our prior work \cite{WAWEnets} was published.
The input to \cite{Jia2020} is a waveform, but the first processing step is an encoder that ``converts the waveform into a feature map
which is similar to the spectrum.'' No magnitude function is applied, so all information is preserved. Experiments in \cite{Jia2020} show the proposal has an advantage over two feature-based approaches when estimating values from two FR tools in the case of speech with added noise.
In \cite{Serra2021} ML is applied to $\mu$-law compressed speech in order to estimate speech quality MOS. But the value of $\mu$ is learned (it is initialized to 256 as in G.711 \cite{G711}), so, in effect, ML is actually applied to the speech waveform, but with the strong suggestion that the first ML step should be compression. No quantization is added, so this approach can preserve all of the information in the speech waveform.

In \cite{Zhang2021} the MOS estimation task is learned in two different ways. Speech waveforms are first processed either by a learnable 1-D convolution layer or by the ``conventional STFT,'' and then supplied to the main ML architecture. Authors report that, compared to the STFT, ``the learnable 1-D convolution layer leads to slight improvements for all targets in nearly all criteria.'' This result suggests that allowing ML to operate on waveforms is preferable to providing it a spectral representation. 

The algorithm presented in \cite{Ryandhimas2023} passes the input speech waveform through a learnable filter bank and combines the resulting information with that from several types of extracted features. The very recently developed estimator in \cite{Kumar2023} uses only the speech waveform as input and we compare its predictions with WAWEnets results in Section \ref{subsec:comparison}. 
In \cite{TorcoliWASPAA21} a WAWEnet-style architecture is applied to waveforms to generate quality information to guide dialog remixing.

In \cite{WAWEnets} we introduce four WAWEnets.  Three of these emulate FR speech quality tools POLQA \cite{POLQA}, WB-PESQ \cite{WBPESQ}, and PEMO \cite{PEMO}. The fourth WAWEnet emulates the FR speech intelligibility tool STOI \cite{STOI}.

Since that time, we have updated the WAWEnet architecture to be more efficient and effective and have leveraged large amounts of data to train three very effective WAWEnets. Section \ref{sec:arch} describes the new architecture. In Section \ref{sec:FR}  we describe how we used 252 hours of speech from 13 languages to train and test individual WAWEnets to very closely emulate seven FR tools (four speech-quality and three speech-intelligibility). For comparison purposes, we provide results from FR versions of these WAWEnets as well. Next we present a single WAWEnet that emulates all seven FR tools at once. 

Section \ref{sec:FR+4subj} introduces 30 hours of subjectively rated speech, and we train another WAWEnet to produce values for four subjective rating dimensions and the seven FR quality and intelligibility values.
In Section \ref{sec:mos} we train and evaluate a third WAWEnet using 52 hours of speech with MOS scores.  This WAWEnet tracks MOS scores quite closely, confirming that the single architecture can be trained to produce very good results for both FR objective targets and subjective targets.

Using ML to develop NR tools for evaluating speech and audio is a rich and active field and numerous other NR tools have been proposed.  Our work is novel because we adopt a relatively simple convolutional architecture, we apply it directly to speech waveforms, we successfully train to closely emulate eleven different targets, and we use an unprecedented quantity and diversity of speech data.
In Section \ref{sec:disc} we provide multiple quantitative comparisons between WAWEnets and some other available NR tools, and we find that WAWEnets compare favorably.
Finally, the homogeneous hierarchical structure of WAWEnets allows us to use the language of signal processing to explain how WAWEnets collapse temporal information into a feature vector that is an evaluation of the original signal. This explanation is provided in Section \ref{spInterp}. To our knowledge, an interpretation of this sort has never before been published.

\begin{figure}
    \centering
    \includegraphics[width=0.8\columnwidth]{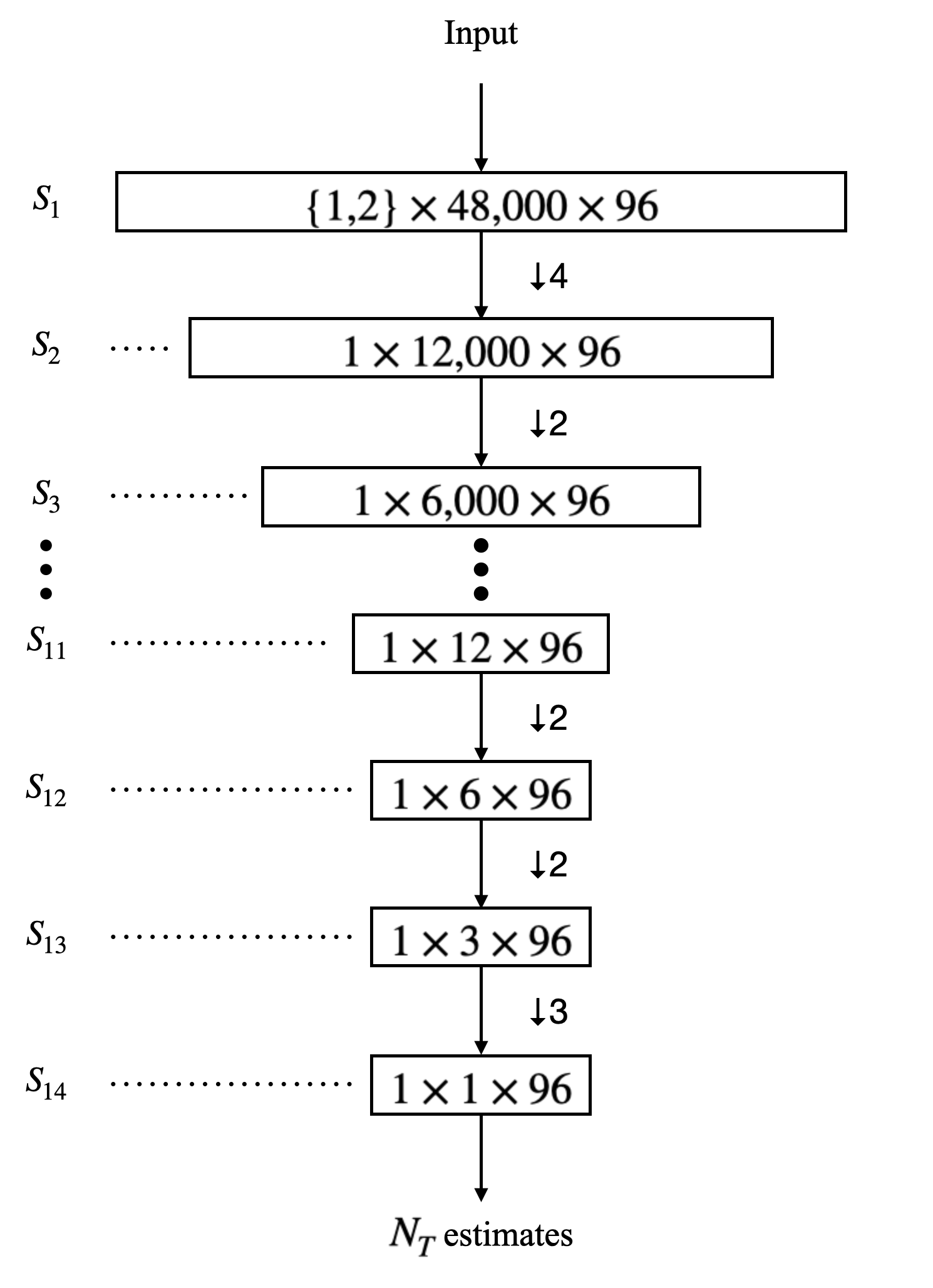}
    \caption{Diagram depicting the overall shape of our WAWEnet model. Each rectangle represents one section, and the enclosed text describes the shape of the section's input vector. The input for all NR WAWEnets is $1\times48,\!000$, allowing only for an impaired speech segment. The input for all FR WAWEnets is $2\times48,\!000$, allowing for an impaired and reference speech segment. Starting with 48,000 speech samples, the model coalesces information in the time domain until a 96-dimension feature vector remains.}
    \label{fig:arch_diagram}
\end{figure}

\section{WAWEnet Architecture}
\label{sec:arch}

WAWEnets use a hierarchy of convolutional layers, pooling, and non-linearities to analyze the quality and intelligibility of one-dimensional wideband speech input signals.
We designed WAWEnets to process wideband speech because wideband speech is quickly becoming the norm in telecommunications. 
The WAWEnets training process optimizes a hierarchy of convolutional filters that can then emphasize or attenuate any frequency in the wideband speech range (nominally 50~Hz--7~kHz) without any prior assumptions or constraints, thus enabling an efficient and effective solution to the task of quality measurements. 
This architecture can also be interpreted using basic signal processing operations as we describe in Section \ref{spInterp}.

Similar to the model we describe in \cite{WAWEnets}, WAWEnets accept 3 seconds of wideband speech sampled at 16,000 samples/second as input. 
The architecture is composed of sections that use one-dimensional, 96-channel convolutional layers to filter the signal, and pooling layers in effect downsample the signal.

This model differs from the model described in \cite{WAWEnets} in a few key ways.
We added additional convolutional sections that downsample all the way to one ``sample'' per channel for each of the 96 channels, thus yielding a 96-dimension feature vector.
A dense layer then maps the feature vector to an estimate.
This model therefore only uses convolutional layers and downsampling to coalesce temporal information, whereas the previous model also used a dense layer to coalesce information across a temporal and a feature dimension.
This new strategy allows the network to gracefully make predictions on segments with significant portions of non-speech data without the need for long short-term memory (LSTM) blocks or other sequence modeling elements. Convolutions near the WAWEnet input may be viewed as optimized filters that operate on speech signals and downsampled speech signals to extract descriptors, while convolutions near the output are more naturally interpreted as processors that optimally combine shorter-term speech signal descriptors to produce longer-term descriptions, ultimately giving a single result for the entire 3-second signal.

All PReLU activations have been replaced with ReLU, and average pooling is used throughout.
The first section downsamples the input signal by four, thereby significantly reducing memory and computation requirements.

In this work we allow the dense layer in $S_{14}$ to generate multiple estimates ($N_T, N_T \geq 1$).
This model is visualized in Fig. \ref{fig:arch_diagram} and is fully described by Tables \ref{table:conv_arch} and \ref{table:conv_blocks}.
With $N_T=1$, the new formulation has a total of 335,905 parameters, a roughly 50\% increase over the 225,025 parameters of our previous model.
However, the number of multiply-accumulates (MAC) has decreased from 968 M to 643 M, a decrease of more than 325 M MAC, or ~34\%.
For brevity, we refer to specific WAWEnet implementations using a subscript that can indicate the type and number ($N_T$) of estimates produced. 
For example, \wawenetmultisubj\ is a WAWEnet with one input channel and 11 estimates.
In addition, \wawenetsubj\ estimates one subjective target and \wawenetobj\ estimates one objective FR target.

We also made a separate WAWEnet configuration that allows $S_1$ to accept two 48,000 sample vectors.
We used this configuration to create an FR WAWEnet that uses reference and impaired speech (\wawenetfr), and, for comparison purposes, a WAWEnet that uses two identical copies of the impaired speech (\wawenettwonr).

\begin{table}[tbh]
\centering
\caption{WAWEnet architecture. Sections $S_1$--$S_{13}$ are composed of one of the two section types listed in Table \ref{table:conv_blocks}. Number of input and output samples per channel are given by $l_{in}$ and $l_{out}$, effective sample rate by $\hat{f_s}$, and effective sample spacing by $s_l$. The dense layer $S_{14}$ maps $f_n=96$ scalar outputs from $S_{13}$ to $N_T$ estimates.}
\begin{tabular}{c|crrcr}
$S$      & section type & \multicolumn{1}{c}{$\hat{f_s}$ (Hz)} & \multicolumn{1}{c}{$l_{in}$} & $s_l$ (ms) & \multicolumn{1}{c}{$l_{out}$} \\ \hline
$S_1$    & Conv A-4   & 16,000                               & 48,000                       & 0.0625     & 12,000                         \\
$S_2$    & Conv A-2   & 4,000                                & 12,000                       & 0.25      & 6,000                         \\
$S_3$    & Conv A-2   & 2,000                                & 6,000                       & 0.5      & 3,000                         \\
$S_4$    & Conv A-4   & 1,000                                & 3,000                        & 1        & 750                         \\
$S_5$    & Conv A-2   & 250                                & 750                        & 4          & 375                           \\
$S_6$    & P Conv A-2   & 125                                  & 376                          & 8          & 188                           \\
$S_7$    & Conv A-2   & 62.50                                 & 188                          & 16         & 94                           \\
$S_8$    & Conv A-2 & 31.25                                 & 94                          & 32         & 47                            \\
$S_9$    & P Conv A-2   & 15.63                                 & 48                           & 64         & 24                            \\
$S_{10}$    & Conv A-2   & 7.81                                 & 24                           & 128         & 12                            \\
$S_{11}$    & Conv A-2   & 3.91                                 & 12                           & 256         & 6                            \\
$S_{12}$    & Conv A-2   & 1.95                                 & 6                           & 512         & 3                            \\
$S_{13}$    & Conv A-3   & 0.98                                 & 3                           & 1024         & 1                            \\
$S_{14}$ & Dense      & —                                    & $f_n$                        & —          & $N_T$                            
\end{tabular}

\label{table:conv_arch}
\end{table}

\begin{table}[tbh]
\centering
\caption{WAWEnet section types. Each section contains a 1-D convolution layer C-$f_n$-$f_l$ with \fn$=96$ filters per channel and \fn\ channels, filter length $f_l=3$, stride of 1, and zero padding $\lfloor\frac{f_l}2\rfloor=1$. Padding layer $\text{Pad}(a, b)$ prepends $a$ zeros and appends $b$ zeros to the input vector. ReLU indicates a ReLU activation. $k$ denotes average pooling layer kernel size.}
\begin{tabular}{c|c c}
Name                     & Conv A-$k$  & P Conv A-$k$ \\ \hline
                         &             & $\text{Pad}(0, 1)$            \\
                         & C-$f_n$-$f_l$   & C-$f_n$-$f_l$      \\
                         & BatchNorm   & BatchNorm      \\
                         & ReLU & ReLU    \\
\multirow{-5}{*}{Layers} & AvgPool-$k$ & AvgPool-$k$   
\end{tabular}

\label{table:conv_blocks}
\end{table}

\section{WAWEnet for Seven Objective FR Targets}
\label{sec:FR}

\subsection{Data}
\label{subsec:data}
The Institute for Telecommunication Sciences (ITS) dataset is formed from high-quality WB speech recordings. We carefully selected these from an array of sources including \cite{Psup23, P.501, TSP, TIMIT, TedNAE, TedSpanish, THCHS30Mandarin, TedKorean, AAFrench, OSR} and also from recordings made in our lab.
We extracted 3-second ``reference segments'' from these recordings such that each reference segment has a speech activity factor (SAF) of 50\% or greater (determined by the P.56 Speech Voltmeter found in \cite{UGST_STL}).
Allowing significant portions of non-speech signal at any location in the segment acts as a natural form of regularization and facilitates WAWEnets' convolutional sections to make quality estimates without the need for explicit sequence modeling elements such as LSTMs.
Any given segment has at most 50\% (1.5 sec) content in common with any other segment. We then normalized each reference segment to have an active speech level of \mbox{$26 \pm 0.2$} dB below the overload point, again using the P.56 Speech Voltmeter.

This process provided 84 hours of speech in the form of 100,681 reference segments, representing 13 languages and 1230 different talkers.
Spanish, Mandarin, and North American English each account for 29\% of the segments. Korean contributes 6\% of the segments, African-accented French 3\%, Japanese 2\%, while Italian, French, German, Portuguese, Hindi, British English, Finnish, Dutch, and Polish combine to contribute the remaining 2\%.

We strategically split the reference segments into training, testing, and validation groups, as well as an unseen group.
We generated the unseen group by reserving 10\% of the talkers in each language, to the extent possible.
The resulting 127 talkers associated with the unseen group do not contribute any segments to the remaining data and are used only for evaluation.
The unseen dataset contains 10,391 segments (9 hours) of speech.
We split the remaining segments into training (50\%), testing (40\%), and validation groups (10\%) with approximate sizes of 38, 30, and 7 hours, respectively.

We processed each reference segment with software to simulate the impairments found in a wide range of current telecommunications environments. These impairments cover three classes:  background noise and suppression, speech codecs, and packet loss and concealment. 

Background noises include coffee shop, party, and street noise at SNRs between 5 and 25 dB. The noise suppression algorithm follows the popular time-frequency (TF) masking paradigm.  It is not intended to be optimal but instead is designed to be easily adjusted to cover a wide operational range, from mild suppression that leaves significant noise and produces minimal artifacts and distortion, to very aggressive suppression that leaves no noise but creates significant artifacts and distortion. To achieve this, the algorithm applies the STFT to each speech signal to produce a TF representation of the signal. It then replaces elements of the TF representation with zero if their magnitude falls below a selected threshold. Finally the inverse STFT and the overlap-and-add (OLA) process convert this new TF representation back to a time-domain speech signal. These conversions are further formalized in \cite{VoranMMSP2021}.

If the threshold is zero, no elements of the TF representation are changed and the original speech signal is reproduced.  Low thresholds remove some elements (typically noise) but not so many elements that artifacts are produced.  Higher thresholds remove more noise and also remove lower level speech components, thus producing more artifacts. We adjusted the threshold from 30 to 60 dB below the level of the peak TF element and also adjusted the STFT window length from 4 to 64 ms duration in order to achieve a wide range of impairment types and levels and thus a wide range of speech qualities and intelligibilities.

We applied 6 WB codec algorithms:  EVS, AMR-WB, Opus, G.711.1, G.722.1, and G.722. We selected bit-rates ranging from 8 to 64 kbps, for a total of 49 WB codec modes. We used 13 different NB codecs, including EVS, AMR, Opus, G.711, G.729, G.723.1, G.726, MELP, and others. Bit-rates ranged from 1.2 to 64  kbps for a total of 40 NB codec modes. We used both independent and bursty packet losses at rates ranging from 5 to 40\%, followed by packet loss concealment (PLC). Finally, we normalized each impaired segment to have an active speech level of \mbox{$26 \pm 0.2$} dB below the overload point.

Each reference segment was impaired in three different ways:  a randomly selected NB noise or codec impairment, a randomly selected WB noise or codec impairment, and a randomly selected NB or WB impairment that combined a codec with noise or PLC or both.  This produced roughly 302,000 segments of impaired speech. This is 252 hours (114 training, 90 testing, 21 validation, and 27 unseen).
A high-level summary of the impairment distribution is given in Table \ref{impairments}. In each row 50\% of the speech is NB and 50\% is WB. A total of 321 distinct conditions are present in the ITS dataset.

\begin{table}[tbh]
\centering
\caption{Distribution of impairments in ITS dataset.}
\label{impairments}
\begin{tabular}{cccc}
\toprule
Hours  & Noise \&  & Speech Codecs &    Packet Loss \&  \\
  & Suppression &  &     Concealment \\
\midrule
$84$    &  \checkmark   &               &           \\
$84$    &               & \checkmark    &           \\
$28$    &  \checkmark   & \checkmark    &           \\
$28$    &               & \checkmark    & \checkmark\\
$28$    &  \checkmark   & \checkmark    & \checkmark\\
\bottomrule
\end{tabular}
\end{table}

Each impaired segment in the dataset was then labeled with values from seven established FR estimators: WB-PESQ \cite{WBPESQ}, POLQA \cite{POLQA}, ViSQOL (compliance version c310) \cite{ViSQOL3}, and PEMO (software available via \cite{PEASS}) estimate the quality of the impaired segment, while STOI \cite{STOI}, ESTOI \cite{ESTOI}, and SIIBGauss \cite{SIIBGauss} give estimates of the speech intelligibility. Each of these seven FR tools compared every impaired segment with the corresponding WB reference segment. We then trained WAWEnets to estimate these FR values using only the impaired segments.

\subsection{Training and Results}
\label{subsec:its_training}
We used affine transformations to map observed values from WB-PESQ ([1.02, 4.64]), POLQA ([1, 4.75]), PEMO ([0, 1]), ViSQOL ([1, 5]), STOI ([0.45, 1]), ESTOI ([0.23, 1]), and SIIBGauss ([0, 750]) to [-1, 1] before use as targets.
As in \cite{catellier2019wenets} and \cite{WAWEnets}, we performed inverse phase augmentation (IPA) to augment all datasets in order to train WAWEnets to learn invariance to waveform phase inversion.
This augmentation increased the amount of data available to just over 500 hours of total speech data.

When training our model from scratch on the ITS dataset, we used one set of initial weights for each training process.
This set of weights was initialized using the Kaiming-Normal initialization method \cite{kaiming_he_norm}.
In the cases where $N_T>1$, the weights in the last layer were duplicated $N_T$ times, resulting in a shape of $N_T\times96$.

We seeded all random number generators such that batch order and batch contents were consistent for every training run.
WAWEnets were trained using mini-batches that were as large as GPU memory would allow; in this case, 60 segments per batch.
We used root mean-squared error (RMSE) as our loss function along with the Adam optimizer \cite{DBLP:journals/corr/KingmaB14} with $10^{-4}$ learning rate, and $L_2$ regularization parameter set to $10^{-5}$.
When the network had trained for an entire epoch, we evaluated the validation set and logged the epoch RMSE loss $E_l$ and per-segment correlation between the target and the WAWEnet output, $\rho_{seg}$.
In the case that $E_l$ on the validation set had not decreased by at least $10^{-4}$ for 5 epochs, we multiplied the learning rate by $10^{-1}$.
The network was trained for 30 epochs.
Training \wawenetobj\ on one NVIDIA GTX 1070 took about 14 hours.\footnote{Certain products are mentioned in this paper to describe the experiment design. The mention of such entities should not be construed as any endorsement, approval, recommendation, prediction of success, or that they are in any way superior to or more noteworthy than similar entities not mentioned.}

WAWEnets are NR tools, but for completeness of the research effort, we also created some FR versions. An NR WAWEnet processes 48,000 samples ($3\text{s} \times 16,000$ samples/second). The FR version (\wawenetfr) processes reference speech and the corresponding impaired speech. The two signals are processed independently at the input layer, and then jointly thereafter. This joint processing allows WAWEnets to compute a family of relevant functions of the two signals. 
To accommodate two input signals, the first convolutional layer of \wawenetfr\ has more parameters (input size $2\ \text{channels}\times48,000\ \text{samples}\times96\ \text{filters}$) compared to the NR WAWEnet (input size $1\ \text{channel}\times48,000\ \text{samples}\times96\ \text{filters}$).
During initialization, all weights in the $S_1$ were duplicated resulting in a shape of $2\times48,000\times96$.
In order to make a fair comparison, we also created a dual input NR version called \wawenettwonr.
\wawenettwonr\ and \wawenetfr\ have identical architecture and number of parameters.
\wawenettwonr\ processes two identical copies of the impaired speech.
Results for all three versions are presented and compared below.

\subsection{Individual network for each of seven FR targets}
We trained individual WAWEnets (\wawenetobj) for each of the seven FR targets. Table \ref{fr_vs_nr_corr} gives the resulting per-segment Pearson correlations and Table \ref{fr_vs_nr_normrmse} shows the corresponding per-segment RMS errors. To allow for direct comparisons, these errors are normalized and shown as a percentage of the full scale for each target.

The network outputs are highly correlated to the FR target values across a vast amount of data that spans a wide range of impairment types, talkers, and languages, in spite of the fact that the networks have no access to the reference speech for comparison purposes.  In effect, the networks embody very effective generalized models for speech quality (or intelligibility).  These generalized models are invariant to speech content, talker, and language, and this allows them to operate without comparison to a reference speech signal.

The ``Dual-NR'' column of Table \ref{fr_vs_nr_corr} shows that the additional weights in \wawenettwonr's $S_1$ have minimal effect on correlation.  And the ``FR'' column (\wawenetfr) shows that access to reference speech (and necessarily increasing the network size) does produce some additional benefit, as expected. Table \ref{fr_vs_nr_normrmse} tells the same story in the RMSE domain.  The networks produce impressive estimation error values that range from 5 to 9\% of full scale. These values barely change when network size is increased, but are further reduced when reference speech is provided. 

\begin{table}[tbh]
\centering
\caption{Pearson correlations between predictions from three WAWEnets with $N_T=1$ and seven objective FR targets, unseen portion of ITS dataset, individual network for each target, correlations calculated per-segment.}
\label{fr_vs_nr_corr}
\begin{tabular}{cccc}
\toprule
Target  & NR & Dual-NR&    FR \\
&\wawenetobj&\wawenettwonr&\wawenetfr\\
\midrule
WB-PESQ       &  0.94 &       0.95 &       0.98 \\
POLQA      &  0.93 &       0.93 &       0.97 \\
ViSQOL     &  0.95 &       0.95 &       0.98 \\
PEMO       &  0.95 &       0.95 &       0.97\\
STOI       &  0.91 &       0.90 &       0.98 \\
ESTOI      &  0.95 &       0.95 &       0.99\\
SIIBGauss  &  0.96 &       0.97 &       0.99\\
\bottomrule
\end{tabular}
\end{table}

\begin{table}[tbh]
\centering
\caption{Normalized RMSE between predictions from three WAWEnets with $N_T=1$ and seven objective FR targets, unseen portion of ITS dataset, individual network for each target, errors calculated per-segment, values are percent of full scale.}
\label{fr_vs_nr_normrmse}
\begin{tabular}{cccc}
\toprule
Target & NR & Dual-NR & FR\\
&\wawenetobj&\wawenettwonr&\wawenetfr\\
\midrule
WB-PESQ       &  9.2 &       9.1 &     4.9\\
POLQA      &  9.4 &       9.4 &     6.6\\
ViSQOL     &  7.2 &       7.0 &     4.9 \\
PEMO       &  6.1 &       6.2 &     5.1\\
STOI       &  5.1 &       5.2 &     2.4\\
ESTOI      &  6.1 &       6.1 &     3.2\\
SIIBGauss &  4.9 &       4.6  &     2.2\\
\bottomrule
\end{tabular}
\end{table}

\subsection{Single network for all seven FR targets}
The outstanding success of individual WAWEnets for each target suggests that the WAWEnet architecture has plenty of capacity for this class of problems. This then suggests the possibility of a single network that maps speech signals to points in a single latent space such that those points can then be mapped to all four quality and all three intelligibility targets. Speech quality indicates how pleasing a speech signal is to the ear and speech intelligibility measures the amount of information carried by the speech. These are certainly different quantities, but they are also related, and this bodes well for the possibility of a single network.

\begin{figure*}

\subfloat[WB-PESQ\label{fig:its_pesqmoslqo}]{\includegraphics[width=44mm]{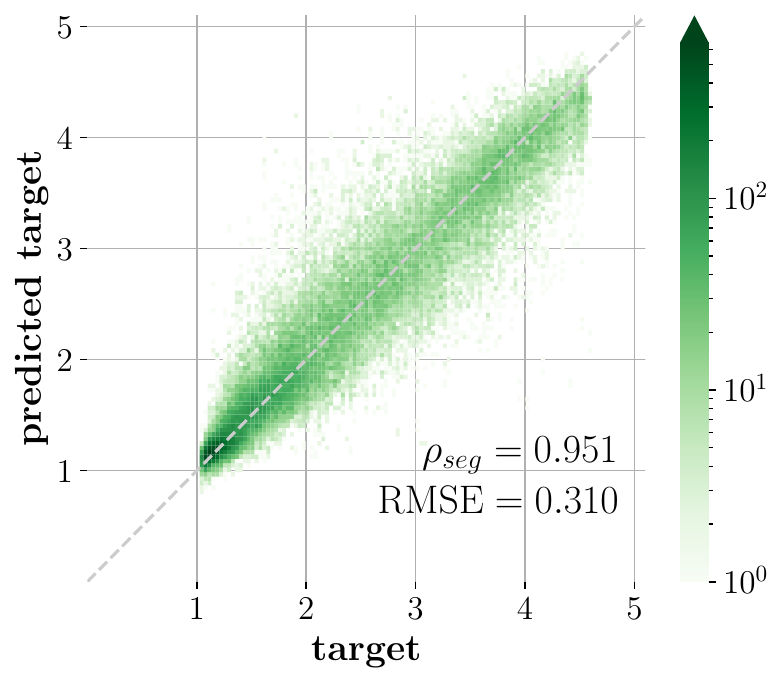}}
\hfill
\subfloat[POLQA\label{fig:its_polqamoslqo}]{\includegraphics[width=44mm]{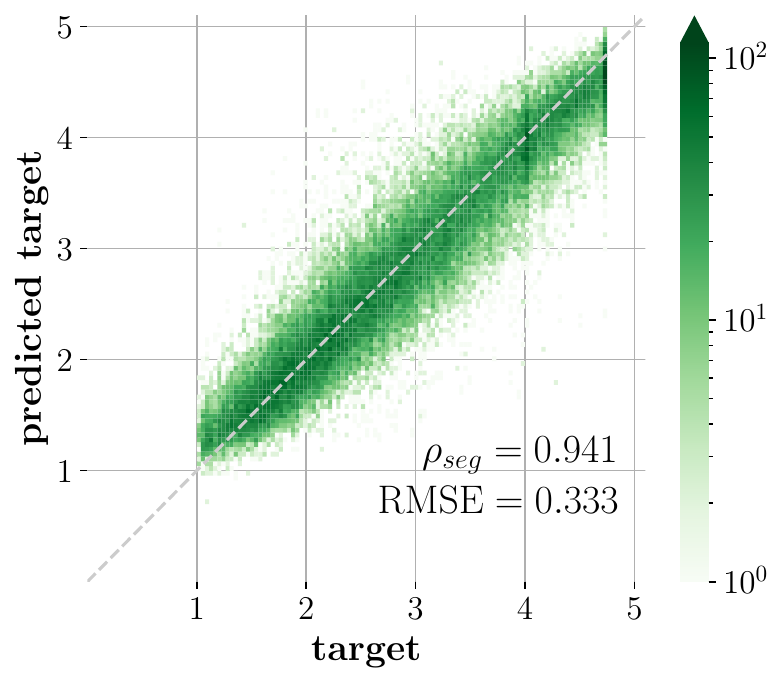}}
\hfill
\subfloat[ViSQOL\label{fig:its_visqol}]{\includegraphics[width=44mm]{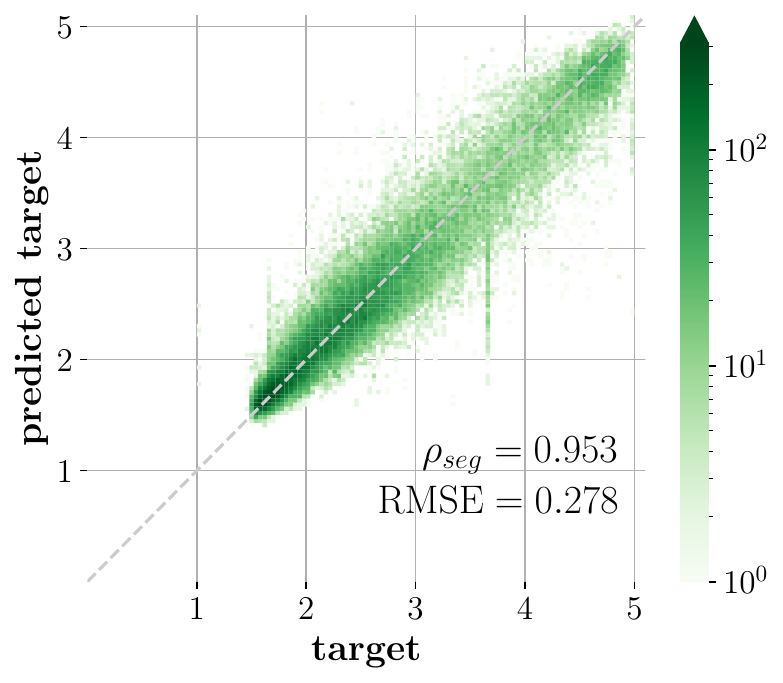}}
\hfill
\subfloat[STOI\label{fig:its_stoi}]{\includegraphics[width=44mm]{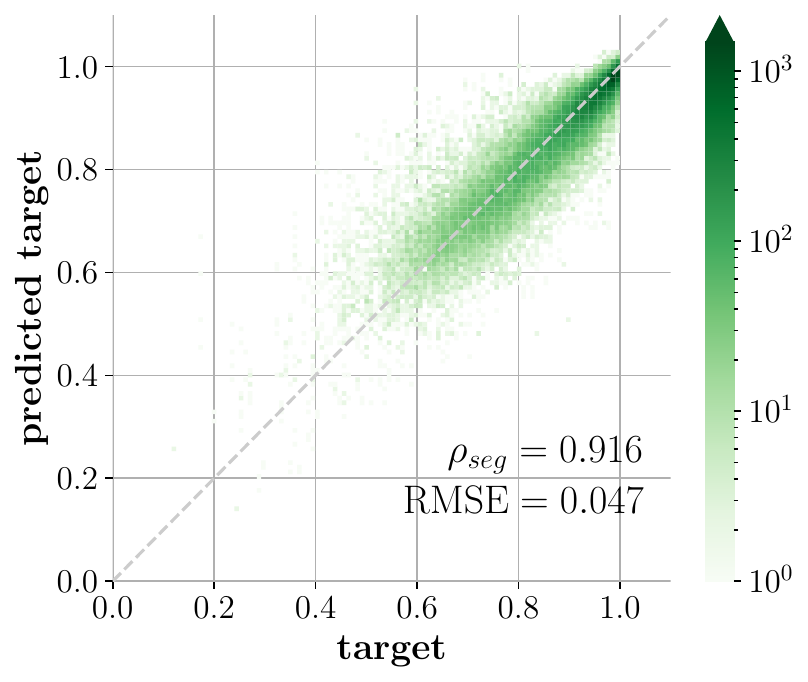}}
\caption{Two-dimensional histogram showing
joint distribution of \wawenetmultiobj\ per-segment estimates and actual target values with per-segment Pearson correlation and unnormalized RMSE for four of the seven targets on the unseen portion of the ITS dataset. Number of segments per bin is given by the scale at the right.}
\label{fig:its_scatter}
\end{figure*}

We trained a single WAWEnet to produce four, five, six, and all seven target values and found that this co-training is both possible and beneficial. Seven networks are replaced with a single network of nearly identical size without compromise in performance. In fact, co-training with seven targets appears to regularize the problem and results in improved performance.  Table \ref{singleNet_7Targets} provides correlation and RMSE values for the single network called \wawenetmultiobj.  The single network correlation values are better than those for the individual networks (see Table \ref{fr_vs_nr_corr}) in four cases and they are matched in the other three cases. The single network RMSE values are better than those for the individual networks (see Table \ref{fr_vs_nr_normrmse}) in six cases and they are matched in the seventh case. Fig. \ref{fig:its_scatter} contains two-dimensional histograms that show the joint distribution of \wawenetmultiobj\ per-segment estimates and actual target values for four of the seven targets.

\begin{table}
\centering
\caption{
Pearson correlation, RMSE, and normalized RMSE between seven estimates from a single network (\wawenetmultiobj) and seven objective FR targets, unseen portion of ITS dataset, correlation and error calculated per-segment.  Final column shows the nominal range for each target as $[\text{val}_{\text{min}}, \text{val}_{\text{max}}]$. Note that $\text{RMSE (\%)}= 100\times\text{RMSE}/(\text{val}_{\text{max}} - \text{val}_{\text{min}})$}.
\label{singleNet_7Targets}
\begin{tabular}{ccccccc}
\toprule
 & Correlation & RMSE & RMSE (\%) & Full Scale \\
\midrule
WB-PESQ  &        0.95  & 0.31 & 7.8 & $[1, 5]$  \\
POLQA &         0.94 & 0.33 & 8.3 & $[1, 5]$ \\
ViSQOL      &     0.95 & 0.28 & 6.9 & $[1, 5]$  \\
PEMO        &     0.96 & 0.06 & 6.1 & $[0, 1]$  \\
STOI        &    0.92 & 0.05 & 4.7 & $[0, 1]$  \\
ESTOI       &   0.95 & 0.06 & 5.8& $[0, 1]$  \\
SIIBGauss   &    0.96 & 35.1 & 4.7 & $[0, 750]$  \\

\bottomrule
\end{tabular}
\end{table}

\section{WAWEnet for Seven Objective FR Targets and Four Subjective Targets}
\label{sec:FR+4subj}
Having successfully developed a single WAWEnet that emulates seven FR tools, we asked if that network might be further trained to also emulate subjective scores.

\subsection{Data}
We added to our collection the dataset described in \cite{Mittag2021IS} and generously provided by the Quality and Usability Lab of the Technische Universit\"at Berlin. We designate this the TUB dataset.
It contains a variety of speech sources, simulated impairments (added background noise, selected codecs, packet loss, bandpass filtering, and clipping) and live impairments (background noise, landline-to-mobile calls, and VoIP calls) Additional details are given in \cite{Mittag2021IS}.
We successfully computed seven FR target values for 14,220 of the TUB speech files.\footnote{The ``TEST\_LIVETALK'' database had no reference files so FR targets could not be calculated. FR estimators occasionally fail to produce valid results so this reduces the usable number of files as well.}
Each of these files also has crowd-sourced subjective ratings of overall speech quality, noisiness, coloration, discontinuity, and loudness.

We used the subjective ratings ``overall speech quality,'' ``noisiness,'' ``coloration,'' and ``discontinuity'' (labeled as MOS, NOI, COL, and DIS, respectively) as targets for WAWEnet training. The ``loudness'' rating is not a practical target for WAWEnets because WAWEnets use normalized input speech levels.  This removes variation in overall signal level, which is a dominating factor behind the ``loudness'' ratings. This normalization could be removed if training for ``loudness'' is desired.

We divided the dataset in two ways. The first was according to the labeling that was provided with it.  We used the 10,903 files (77\%) from ``TRAIN\_LIVE'' and ``TRAIN\_SIM'' for training, the 642 (4\%) files from ``TEST\_FOR,'' ``TEST\_NSC,'' and ``TEST\_P501'' for testing, and the 2,675 files (19\%) from ``VAL\_LIVE'' and ``VAL\_SIM'' for validation.
Note that ``TEST\_NSC'' contains German language speech and the remainder of the dataset is English language speech. 
In order to compare this somewhat heterogeneous division with a more homogeneous division, we also divided the dataset through random sampling: 50\% (7,110 files) were used for training, 40\% (5,688 files) for testing, and 10\% (1,422 files) for validation.
The results presented below are based on the testing portion in both cases.

File lengths range from 4.5~s to 14.6~s with a mean of 8.8~s and a median of 9.0~s.  WAWEnets work on 3-second segments where the SAF is at least 50\%. For each file we find all such distinct segments in the file---98\% of the files produce two segments, 48\% produce three and 10\% produce four. The result is approximately 28,200 training segments, 6,800 validation segments, and 1400 testing segments---a total of 30 hours of speech.
Using G.191 tools \cite{UGST_STL}, we converted the data from 48,000~samples/second to 16,000~samples/second and normalized each segment to \mbox{$26 \pm 0.2$}~dB below the overload point.

Each subjective rating is based on an entire file. For training, we replicate that file rating to create an identical target for each segment extracted from that file.  If the file shows little temporal variation in the rated attribute, then this target replication incurs only a minor approximation.  But if there is major variation (e.g. localized packet loss or non-stationary background noise), then replication of targets can be a significant approximation and source of error. 
For testing, the correlations and RMSE values compare each per-segment WAWEnet output with the subjective rating of the corresponding file.

Signal bandwidths create an additional approximation in this work. The subjective ratings in the TUB dataset are based on FB speech signals but WAWEnets are WB and only analyze the lower 8 kHz of these signals. There is no principled method to convert FB MOS values to WB MOS values (this is connected with the fact that MOS values are relative, not absolute) so we use the FB values as is and accept the approximation.  The spectrum above the WB upper limit typically makes relatively small contributions to quality and intelligibility so we consider this to be a very close approximation.

\begin{figure*}[h]
\subfloat[MOS\label{fig:tub_504010_mos}]{\includegraphics[width=44mm]{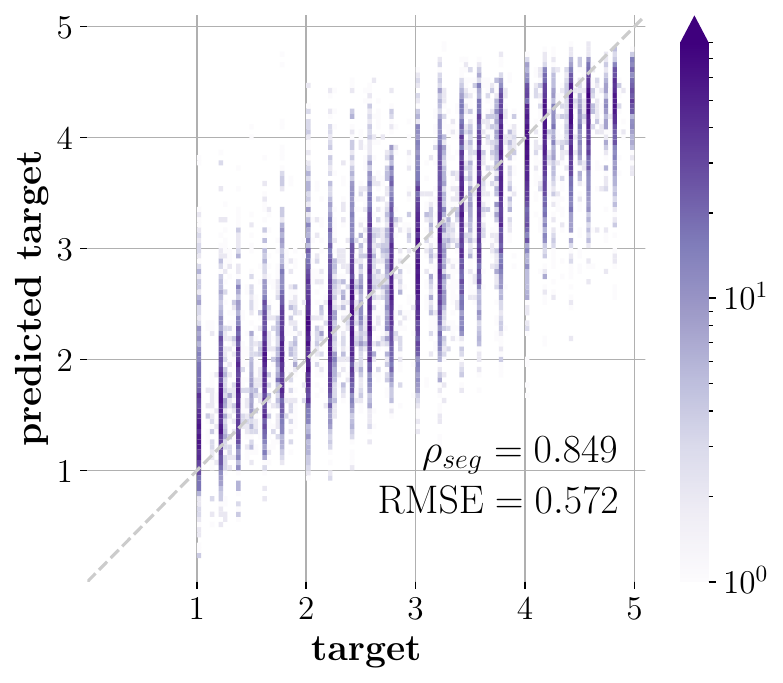}}
\hfill
\subfloat[DIS\label{fig:tub_504010_dis}]{\includegraphics[width=44mm]{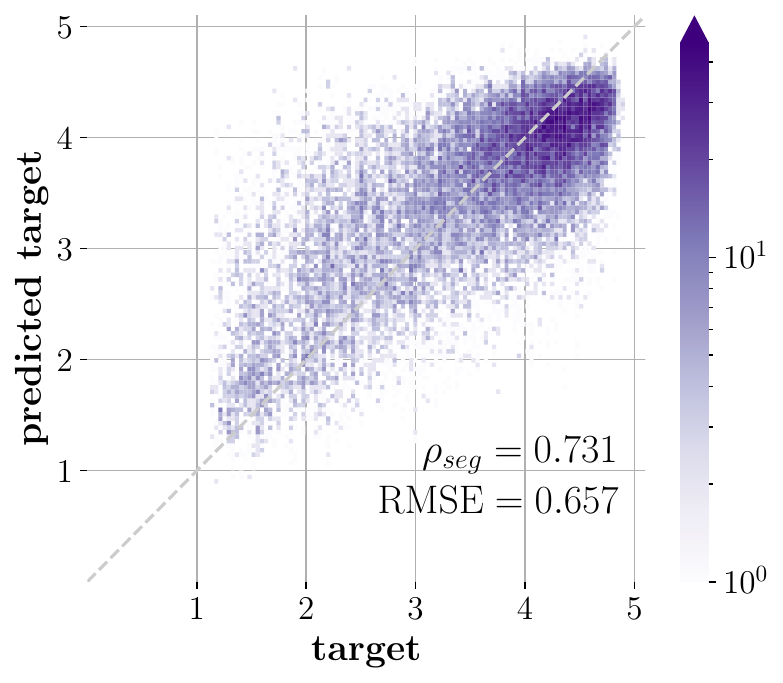}}
\hfill
\subfloat[WB-PESQ\label{fig:tub_504010_pesq}]{\includegraphics[width=44mm]{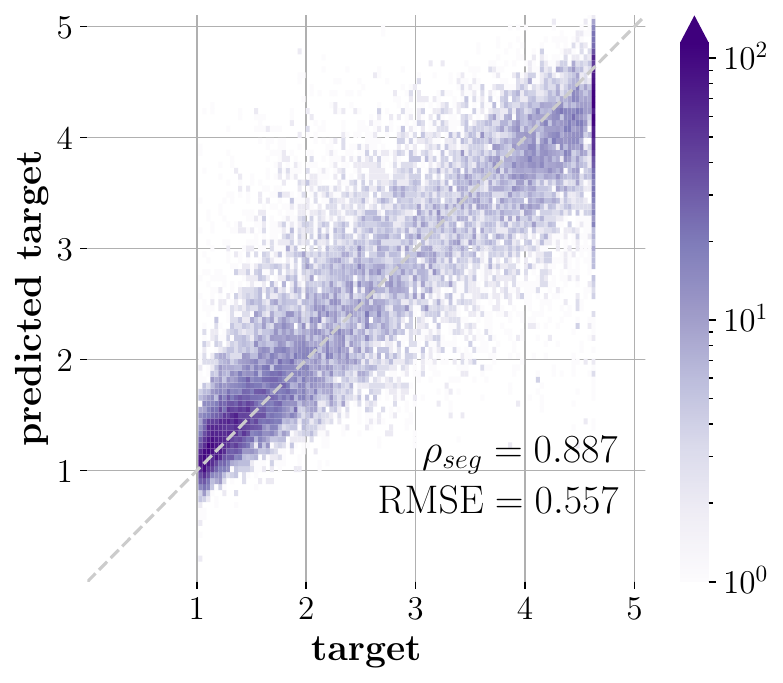}}
\hfill
\subfloat[STOI\label{fig:tub_504010_stoi}]{\includegraphics[width=44mm]{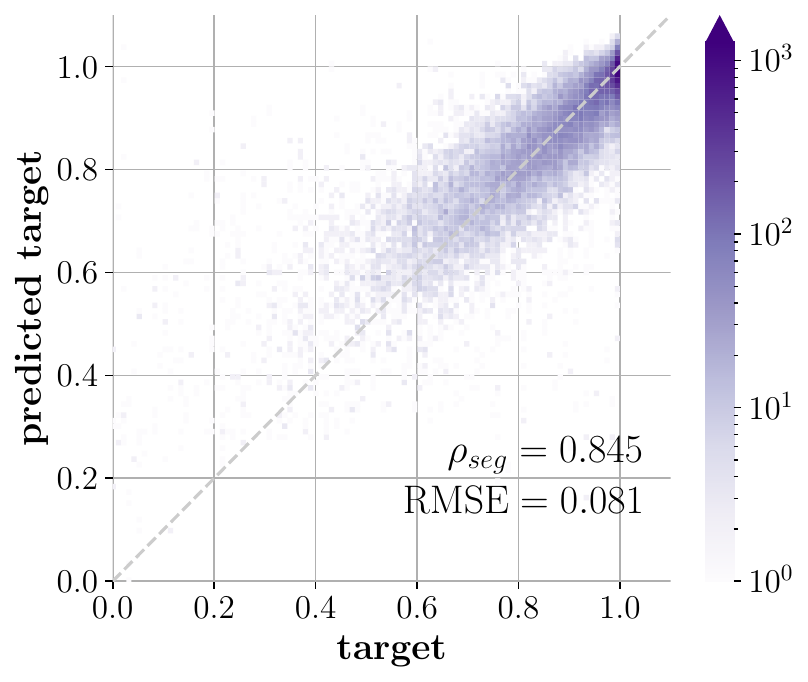}}
\caption{Two-dimensional histogram showing joint distribution of \wawenetmultisubj\ per-segment estimates and actual target values with per-segment Pearson correlation and unnormalized RMSE for four of the eleven targets on the test portion of the TUB dataset when using the 50/40/10 split. Number of segments per bin is given by the scale at the right.}
\label{fig:tub_504010_scatter}
\end{figure*}

\subsection{Training and Results}
Starting with the weights from \wawenetmultiobj\,we allowed the optimizer to update the weights in each section.
This strategy improved overall performance compared to the random initialization strategy used in Section \ref{subsec:its_training}.
We used affine transformations to map all subjective scores from [1, 5] to [-1, 1].
Besides those exceptions, we followed the training process described in Section \ref{subsec:its_training}.

We trained a WAWEnet to emulate the four subjective ratings: \wawenetfoursubj. The per-segment correlation and normalized RMSE values are shown in Table \ref{NISQA_res_per_file}.  We also trained WAWEnets to emulate the four subjective ratings and three, four, five, six, or all seven of the FR values.
Table \ref{NISQA_res_per_file} also shows the result for a single WAWEnet that emulates four subjective and seven objective FR targets: \wawenetmultisubj.
Fig.~\ref{fig:tub_504010_scatter} contains two-dimensional histograms that show the joint distribution of \wawenetmultisubj\ per-segment estimates and actual target values for four of the eleven targets using the 50/40/10 split.

Table \ref{NISQA_res_per_file} makes clear that co-training with the objective FR targets improves correlations and RMS errors for the four subjective targets. It appears that the extra constraints regularize the problem and lead to a better solution.
Considering the test data prescribed by the TUB dataset, Table \ref{NISQA_res_per_file}  shows dramatic correlation drops and RMS error increases for the FR targets compared to those reported in the previous section.
But the correlation drop is smaller when evaluated on the random split.
Table \ref{singleNet_7Targets} and Table \ref{NISQA_res_per_file} are based on different and dissimilar datasets, so the comparison is not exact.

Estimators are often judged by per-condition statistics.  Target values are averaged for all results from each condition (e.g., each individual codec mode or noise environment) and the same is done for the estimates. These per-condition averages are then compared by correlation or error statistics.
Averaging within each condition reflects a common and relevant estimation situation: it removes variation due to talkers, utterances, and other factors so we can draw clear conclusions about the conditions or about the systems that we are testing.
Table \ref{NISQA_res_per_cond} provides per-condition results analogous to those in Table \ref{NISQA_res_per_file}. Here we see that \wawenetmultisubj\ gives per-condition correlations of 0.85 or better for four subjective and three objective FR targets.
For some targets, \wawenetmultisubj\ performs much better under the  50/40/10 split than under the 77/4/19 split. This indicates that testing on just 4\% of the data (642 files) does not lead to robust or representative results, especially when those 642 files show characteristics that differ from those of the training data, as indicated in the dataset's documentation.

\begin{table}
\centering
\caption{Per-segment Pearson correlation (\rhoseg) and normalized RMSE between single WAWEnet trained for four (\wawenetfoursubj) or eleven (\wawenetmultisubj) targets. Testing portion of TUB dataset. Error values are percent of full scale.}
\label{NISQA_res_per_file}
\begin{tabular}{c|cccc|cccc}
\toprule
{} & \multicolumn{4}{c}{77/4/19 Split} & \multicolumn{4}{c}{50/40/10 Split}\\
{} & \multicolumn{2}{c}{\rhoseg} & \multicolumn{2}{c}{RMSE (\%)}&\multicolumn{2}{c}{\rhoseg} & \multicolumn{2}{c}{RMSE (\%)}\\
$N_T$ &          4  & 11 & 4 & 11&4&11&4&11\\
\midrule
MOS       &       0.80  & 0.82 & 16 & 15&0.84&0.85&15&14\\
NOI       &       0.80  & 0.82  & 14 & 13 &0.80&0.82&14&14\\
COL       &       0.79  & 0.81  & 13 & 12&0.78&0.80&15&14\\
DIS       &       0.75  & 0.78  & 17 & 16&0.71&0.73&17&16\\
WB-PESQ      &           - & 0.79  & - & 19&-&0.89&-&15\\
POLQA     &           - &  0.80  & -  & 17&-&0.88&-&14\\
ViSQOL3   &           - &  0.83  & -  & 14&-&0.91&-&12\\
PEMO      &           - &  0.58  & - & 17&-&0.85&-&8\\
STOI      &           - &  0.52  & - & 16&-&0.85&-&8\\
ESTOI     &           - & 0.58  & - & 22&-&0.88&-&10\\
SIIBGauss &           - &  0.75  & - & 16&-&0.82&-&19\\
\bottomrule
\end{tabular}
\end{table}

\begin{table}
\centering
\caption{Per-condition Pearson correlation ($\rho$) and normalized RMSE between single WAWEnet trained for 4 (\wawenetfoursubj) or 11 (\wawenetmultisubj) targets. Testing portion of TUB dataset. Error values are percent of full scale.}
\label{NISQA_res_per_cond}
\begin{tabular}{c|cccc|cccc}
\toprule
{} & \multicolumn{4}{c}{77/4/19 Split} & \multicolumn{4}{c}{50/40/10 Split}\\
{} & \multicolumn{2}{c}{$\rho$} & \multicolumn{2}{c}{RMSE (\%)}&\multicolumn{2}{c}{$\rho$} & \multicolumn{2}{c}{RMSE (\%)}\\
$N_T$ &          4  & 11 & 4 & 11&4&11&4&11\\
\midrule
MOS & 0.91  & 0.91  & 12 & 12&0.96&0.96&9&9\\
NOI       & 0.91  & 0.91  & 10 &  9 &0.97&0.97&9&9\\
COL      & 0.91  & 0.91  &  9 &  9&0.97&0.97&7&8\\
DIS   & 0.90  & 0.90  & 12 & 12&0.97&0.96&10&10\\
WB-PESQ            &   -   & 0.85  & -  & 15&-&0.94&-&10\\
POLQA           &   -   & 0.91  & -  & 10&-&0.96&-&10\\
ViSQOL3         &   -   & 0.89  & -  & 12&-&0.97&-&8\\
PEMO            &   -   & 0.66  & -  & 18&-&0.93&-&13\\
STOI            &   -   & 0.66  & -  & 19&-&0.95&-&11\\
ESTOI           &   -   & 0.60  & -  & 25&-&0.91&-&15\\
SIIBGauss       &   -   & 0.80  & -  & 13&-&0.88&-&10\\
\bottomrule
\end{tabular}
\end{table}

\section{WAWEnet for Subjective Scores Only}
\label{sec:mos}
We also trained a WAWEnet to closely emulate subjective speech quality scores only.

\subsection{Data}
We are very grateful that the Audio, Speech, and Information Retrieval Group at Indiana University Bloomington provided us with the dataset described in \cite{Dong2020}. We call this the IUB dataset, and it is better suited to WAWEnets than the TUB dataset is.  The IUB dataset is WB, so WAWEnets need not approximate FB subjective scores from WB signals, as was the case with the TUB dataset.  In addition, the IUB file lengths range from  2.0 s to 7.8 s with a mean of 3.8 s and a median of 3.7 s, thus providing a much closer match to the WAWEnet 3-second window than was possible with the TUB data.

The IUB dataset includes high-quality speech from close-talking microphones and lower quality speech from more distant microphones. The distant microphones necessarily capture more natural and artificial environmental noise (SNRs reported in the range $-10$ to +11 dB) and natural reverberation (speech-to-reverberation ratios reported in the range $-5$ to +4 dB)\cite{Dong2020}. In addition, some recordings were subjected to 3.4 kHz low-pass filtering in order to create anchor conditions for subjective testing. Subjective testing was crowd-sourced. Scores were collected, filtered, and normalized \cite{Dong2020} to produce speech quality scores on a scale of zero to 10. The result is 36,000 speech files, each with a scaled subjective speech quality MOS value.

The IUB dataset contains 35,428 files that are 6 seconds or shorter.
From each of these files we selected all disjoint 3-second segments with SAF at or above 50\%. Any file shorter than 3 seconds was zero-padded to create a single 3-second segment. By this process, 1,794 files produced two segments each and the remaining 27,572 files produced one segment each.  This gives more than 31,000 segments, or 26 hours, of speech data.
Each segment was assigned the scaled MOS value of the file that it came from. Using only files of 6 seconds or shorter means that a 3-second segment contains at least half of the file. This minimizes error associated with assigning per-file MOS values to individual segments when speech quality is not constant throughout the file.
Using G.191 tools \cite{UGST_STL}, we normalized each file to \mbox{$26 \pm 0.2$} dB below the overload point.

\subsection{Training and Results}
\label{subsec:IUBTrainingAndResults}
We used affine transformations to map all subjective scores from [0, 10] to [-1, 1].
Starting with the weights from \wawenetmultiobj\, we allowed the optimizer to update the weights in each section.
However, with this dataset equivalent results are achieved using the initialization method described in Section \ref{subsec:its_training}.
We randomly selected 50\% of the segments for training, 10\% for validation and 40\% for testing. 
The rest of the training process was similar to the process described in Section \ref{subsec:its_training}, except we allowed training for 60 epochs in this case.
The results that follow are based on the approximately 14,700 testing segments, which comprise about 12 hours of speech.

\begin{figure}
   \centering
   \includegraphics[width=85mm]{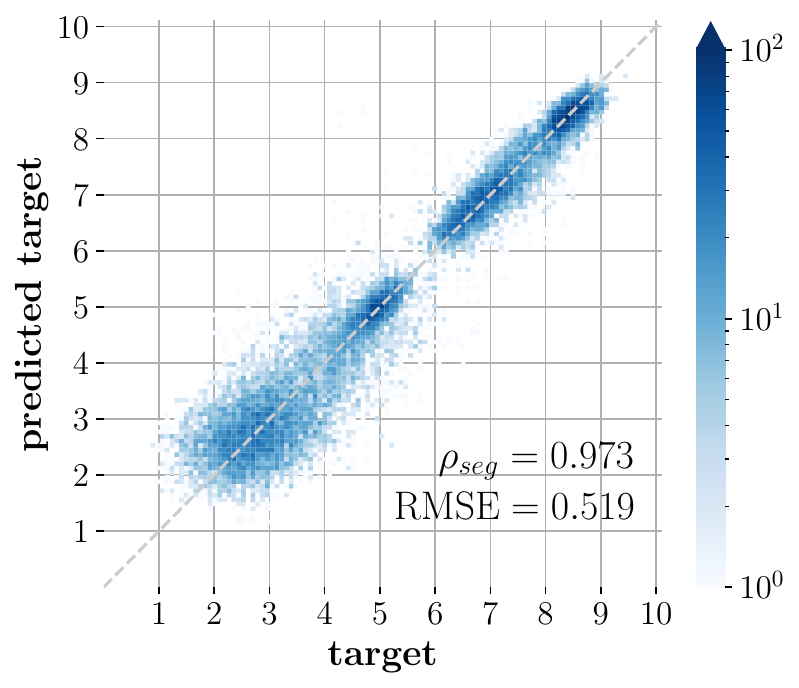}
  \caption{Two-dimensional histogram showing joint distribution of \wawenetsubj\ per-segment estimates and actual target values  with per-segment Pearson correlation and unnormalized RMSE for scaled MOS on the test portion of the IUB dataset. Number of segments per bin is given by the scale at the right.}
   \label{fig:iub_scaled_mos}
\end{figure}

\wawenetsubj\ achieves a per-segment correlation to MOS of 0.97 and normalized RMSE of 5.8\% of full scale.
Fig. \ref{fig:iub_scaled_mos} shows this correlation graphically.
Six other recently proposed NR tools are applied to the IUB dataset in \cite{Zhang2021}, and the resulting Pearson correlations to MOS range from 0.93 to 0.96. The \wawenetsubj\ MOS correlation of 0.973 improves upon the best of these. Mean absolute errors (MAEs) given in \cite{Zhang2021} range from 0.40 to 0.50 and the \wawenetsubj\ MAE is 0.37, which places \wawenetsubj\ at a lower error rate than the best previous result.  
Note that this is not a complete comparison because the best tools in \cite{Zhang2021} produce MOS and three additional estimates, while \wawenetsubj\ produces only MOS.
But there is no question that this WAWEnet architecture performs on a par with these other top performers on this dataset.

\section{Discussion and Comparisons}
\label{sec:disc}

We have shown that the NR WAWEnet architecture can evaluate wideband speech signals in a manner consistent with a variety of objective and subjective evaluations without using any reference speech signal.
\wawenetmultiobj\ can simultaneously very closely emulate seven diverse objective targets. 
\wawenetsubj\ closely emulates MOS values.
And simultaneous emulation of seven objective targets and four subjective targets with \wawenetmultisubj\ appears to improve performance on subjective targets compared to emulating solely subjective targets.

The weights present in \wawenetmultiobj\ appear to be relevant to both the TUB and the IUB datasets.
Training on the TUB dataset with the 50/40/10 split with a random initialization and eleven targets resulted in MOS, NOI, COL, and DIS per-segment correlations of 0.80, 0.80, 0.75, and 0.64, respectively---lower than the correlations achieved when starting from \wawenetmultiobj.
When training on the IUB dataset, the weights present in \wawenetmultiobj\ don't improve estimation performance, but the training process converged to the best performance roughly 5 epochs sooner.
With both the TUB and the IUB datasets, solely allowing the weights in \wawenetmultiobj\ $S_{14}$ to be trained resulted in worse performance than allowing the weights in all sections to be trained.

If desired, one could use WAWEnets to process narrowband speech ($f_s=8000$ samples/second) by modifying input section $S_1$ to have pooling layer size $k=2$ instead of $k=4$.
It is possible to extend the WAWEnet architecture to wider bandwidths (higher sample rates) by adding one or more sections to the input of the network.
For example, adding a section with a $k=3$ would allow WAWEnets to process 3 seconds of speech with a sample rate of 48 kHz.

Likewise, there are several strategies suitable for extending WAWEnets to process longer signals.
One strategy would be to insert a section with a $k=2$ between $S_{13}$ and $S_{14}$.
This would allow WAWEnets to generate estimates for 6-second signals.
Another strategy would be to use either the 96-D feature vector or target estimates as an input to a recurrent neural network of some kind, e.g., LSTM or gated recurrent unit (GRU), thus allowing WAWEnets to process arbitrary-length speech.
With sufficient data, either strategy might learn to properly account for the various principles at work when speech quality varies. These principles include (see \cite{voran_mq05}): long-term ratings are lower-bounded by minima and upper-bounded by averages, larger and more frequent quality variations reduce quality, and recency. Alternatively, longer signals can be accommodated by a sliding 3-second WAWEnet processing window followed by averaging or more sophisticated processing of the multiple results produced.

\begin{table}[!t]
\centering
\caption{
Number of parameters, number of parameters trained, and multiply-accumulates (MAC) per 3 seconds of speech for each of the systems compared in Section \ref{subsec:comparison}.}
\label{table:net_meta}
\begin{tabular}{l|rrr}
\toprule
 & Total Params & Params Trained & MAC \\
 \midrule
NISQA-DIM & 269,486 & 269,486 & 369M \\
\wawenetmultiobj & 335,905 & 335,905 & 643M \\
& & &  \\
wav2vec frozen & 94,372,481 & 769 & 19,490M \\
wav2vec fine-tuned & 94,372,481 & 94,372,481 & 19,490M \\
TorchAudio-Squim & 7,390,000 & 0 & 24,140M\\
\bottomrule
\end{tabular}
\end{table}

\begin{table*}[!t]
\centering
\caption{
Per-segment correlation (\rhoseg) and RMSE (\%) achieved by select models that were trained or fine-tuned on the training portion of the ITS dataset and/or evaluated on the unseen portion of the ITS dataset. Models above the blank rows were trained from scratch on the ITS dataset and models below the blank rows were either fine-tuned on the ITS dataset or not retrained. Please refer to Table \ref{table:net_meta} for information about number of model parameters for each model and number of parameters trained.}
\label{table:comparison}
\begin{tabular}{l|rrrrrrr|c}
\toprule
 & \multicolumn{7}{c|}{\rhoseg} &  \\
 & WB-PESQ & POLQA & ViSQOL & PEMO & STOI & ESTOI & SIIBGauss & mean \\
\arrayrulecolor{black}\hline
NISQA-DIM & 0.960 & 0.949 & 0.947 & 0.934 & 0.915 & 0.950 & 0.971 & 0.947 \\
\wawenetmultiobj & 0.951 & 0.941 & 0.953 & 0.956 & 0.916 & 0.952 & 0.963 & 0.947 \\
& & & & & & & & \\
wav2vec frozen & 0.862 & 0.866 & 0.890 & 0.888 & 0.838 & 0.904 & 0.900 & 0.878 \\
wav2vec fine-tuned & \textbf{0.972} & \textbf{0.961} & \textbf{0.969} & \textbf{0.967} & \textbf{0.946} & \textbf{0.970} & \textbf{0.984} & \textbf{0.967} \\
TorchAudio-Squim &  0.742 & - & - & - & 0.769 & - & - & 0.755 \\ 
\arrayrulecolor{black}\midrule
{} & \multicolumn{7}{c|}{RMSE (\%)} &  \\
 & WB-PESQ & POLQA & ViSQOL & PEMO & STOI & ESTOI & SIIBGauss & mean \\
\hline
NISQA-DIM & 7.2 & 7.9 & 8.0 & 8.3 & 4.7 & 5.9 & 4.6 & 6.7 \\
\wawenetmultiobj & 7.8 & 8.3 & 6.9 & 6.1 & 4.7 & 5.8 & 4.7 & 6.3 \\
& & & & & & & & \\
wav2vec frozen & 13.2 & 12.5 & 10.7 & 9.8 & 6.7 & 8.6 & 7.6 & 9.9 \\
wav2vec fine-tuned & \textbf{6.1} & \textbf{7.1} & \textbf{5.8} & \textbf{5.5} & \textbf{4.1} & \textbf{4.9} & \textbf{3.3} & \textbf{5.3} \\
TorchAudio-Squim &  16.9 & - & - & - & 8.0 & - & - & 12.5 \\
\bottomrule
\end{tabular}
\end{table*}

\subsection{Comparisons}
\label{subsec:comparison}

Part of evaluating model performance is making comparisons to other models that perform the same task.
Care must be taken to ensure equal footing is given in each comparison, and it is often difficult to make perfectly equitable comparisons.
For example, some models in the literature are trained using fully supervised approaches on datasets with impairments that encompass a narrow scope, i.e., only reverb, only noise and noise suppression, or only speech synthesis.
Others are trained using unsupervised or self-supervised approaches on much larger datasets.
Additionally, models vary greatly in size (number of parameters) and computational complexity.

In Section \ref{subsec:IUBTrainingAndResults} we have already reported that the \wawenetsubj\ correlation to MOS of 0.97 on the IUB data is better than any of the six alternatives studied in \cite{Zhang2021}  (correlations 0.93 to 0.96). We will now discuss comparisons we have conducted using freely available models and training code.
In order to make additional fair and informative comparisons, we have selected a few diverse models and training strategies.

NISQA-DIM \cite{Mittag2021IS} is a model that is used to predict MOS and other speech qualities on fullband data. It uses calculated features---specifically a log-mel-spectrogram with 48 bands)---as inputs. We have adapted its open-source training code\footnote{https://github.com/gabrielmittag/NISQA} to predict seven targets, trained it from scratch on our dataset for 30 epochs, and evaluated it on our unseen dataset.

Wav2vec 2.0 \cite{wav2vec} was originally trained in a self-supervised manner for the task of speech recognition on a very large corpus of data. It uses wideband waveforms as inputs and is a very large model.
Borrowing from the approach described in \cite{CooperICASSP2022},\footnote{https://github.com/nii-yamagishilab/mos-finetune-ssl} we used the features generated by the small wav2vec 2.0 model with no fine-tuning\footnote{https://github.com/facebookresearch/fairseq/tree/main/examples/wav2vec} as an input to a linear transformation from 768 features to seven target values. We trained the model in two ways: allowing all model parameters to be updated (wav2vec fine-tuned), and only training the linear transformation (wav2vec frozen). These models were used in place of WAWEnets in our training code, and therefore the same training and testing procedure used for training WAWEnets was implemented.

TorchAudio-Squim \cite{Kumar2023} is large, very computationally complex model that was trained specifically to predict objective (WB-PESQ, STOI) and subjective (MOS) speech quality measurements. It also utilizes wideband waveforms as inputs. We evaluated the models as-delivered in the \texttt{torchaudio} package on our unseen dataset as intended by the authors.\footnote{The model used in this test was gathered from the Python `torchaudio' package, version \texttt{2.1.0.dev20230516+cu118}.} In order to make MOS predictions, a ``non-matching reference'' is required, and we selected a clean source file from the IUB dataset.

A comparison of model characteristics is shown in Table \ref{table:net_meta}. Note that NISQA-DIM and \wawenetmultiobj~have similar size and MAC counts while the other options are much larger and require far more computations. Corresponding evaluation results for estimating FR objective targets are provided in Table \ref{table:comparison}.
When allowed to update all parameters during the training process, the wav2vec model achieves the best performance.
The value of the unsupervised pre-training on a very large dataset is evident, as is the ease of supervised fine-tuning to perform a task that differs greatly from the model's original intended task. When we train only the linear transformation that follows wav2vec, we observe fairly good performance, but this also demonstrates that the latent space generated by the original wav2vec network is not perfectly suited for the purpose of predicting speech qualities.

Setting aside the much larger and more computationally demanding wav2vec and TorchAudio-Squim, we focus now on the much more comparable NISQA-DIM and \wawenetmultiobj.
\wawenetmultiobj~correlations are better than NISQA-DIM correlations for four of the seven targets.
\wawenetmultiobj~RMSEs are better than NISQA-DIM RMSEs for three of the targets and matched for a fourth.
When performance is averaged across all seven targets, the two are matched in correlation, and \wawenetmultiobj~shows a small advantage in RMSE. 
In the default configuration, NISQA-DIM uses a self-attention mechanism and attention pooling in its network architecture.
This enables impressive performance with fewer parameters than \wawenetmultiobj~, and the NISQA-DIM network itself performs fewer MACs than \wawenetmultiobj. Note, however, that computation of the input log-mel-spectrograms required by NISQA-DIM is not included in this computational complexity measurement.

TorchAudio-Squim performed reasonably well when estimating WB-PESQ and STOI on the unseen portion of the ITS dataset, especially in consideration of the comparably narrow scope of its training data.
We also used TorchAudio-Squim to estimate the MOS values on the test portion of the IUB dataset and on the test portion of the TUB dataset with the 50/40/10 split.
Per-segment correlations were 0.417 and 0.197, respectively, and normalized RMSE values were 28.0\% and 37.5\%.
The low correlation and high RMSE indicate poor agreement with MOS values for these two datasets.
These results demonstrate that care must be taken even when using large and thoroughly trained models to evaluate data that may be dissimilar to the model's training data.

These comparisons demonstrate that WAWEnets provide a viable and competitive approach to predicting speech qualities while maintaining accuracy and ease of training. We have shown that training on relatively small datasets produces good results. WAWEnets' small computation and storage footprint minimizes their power consumption and facilitates their use in a wide range of applications. Further, the homogeneous and hierarchical structure of WAWEnets makes them amenable to some level of interpretability. In the next section we use the language of signal processing to describe the internal operation of WAWEnets.

\section{Signal Processing Interpretation}
\label{spInterp}
We have established that the WAWEnet architecture offers an efficient and effective tool for evaluating wideband speech waveforms.  It does this by mapping a wideband speech waveform to a 96-dimensional vector in a latent space. Different projections in that space produce scalar values that can track different objective or subjective values related to speech quality or intelligibility.
We also seek to understand, to the extent possible, \emph{how} the WAWEnet architecture maps waveforms to this space. As is often the case with algorithms developed through machine learning, a fully satisfying interpretation is elusive.
But we can describe the signal processing that a WAWEnet applies and we can give a high-level description of how this signal processing converts a waveform into an evaluation of that waveform.

\subsection{Functions}
To map waveforms to the latent space, a WAWEnet uses 13 sections and each section consists of four layers. In the language of ML, these four layers are convolution, batch normalization, ReLU, and average pooling.  Table \ref{table:spFunctions} shows how the four layers map to six signal-processing (SP) functions.  
Linear time-invariant (LTI) systems are often relatively amenable to analysis.  Table \ref{table:spFunctions} shows that WAWEnets are linear except for the bias and half-wave rectification (HWR). Of course these non-linear functions are what enable WAWEnets (and many other ML based algorithms) to accomplish the assigned tasks and also prevent the network from simplifying into a trivial and ineffective one. Time-invariance is satisfied by all functions except the sub-sampling (which shows time-invariance only for time shifts that are integral multiples of the output sampling period).

\begin{table}[t!]
\centering
\caption{
WAWEnet ML layers expanded into signal processing functions.}
\label{table:spFunctions}
\begin{tabular}{p{.7cm} p{1.9cm} p{2.39cm} p{.5cm} p{.5cm}}
\toprule
ML Layer &  SP Function &    Equation & Linear &Time-Invar\\

\midrule
Conv.       &     FIR Filtering & $y_k = \sum\limits_{i=0}^{2} h_i x_{k-i}$             & ~~\checkmark  & \checkmark \\
\arrayrulecolor{gray}\midrule
\multirow{2}{*}{\parbox{1cm}{Batch Norm.}}  &     Gain          &    $y_k = ax_k$                                       & ~~\checkmark  & \checkmark  \\
                 &       Bias        &     $y_k = x_k + b$                                   &             & \checkmark \\
 \midrule
ReLU        &   Half-Wave Rect. &   $y_k = \text{max}(x_k,0)$                           &             & \checkmark \\
\midrule
\multirow{2}{*}{\parbox{1cm}{Avg. Pool}}    &    FIR Filtering  &   $y_k = \frac{1}{m}\sum\limits_{i=0}^{m-1}x_{k-i}$   & ~~\checkmark  & \checkmark \\
                &     Sub-sampling  &   $y_k = x_{mk}$                                        & ~~\checkmark  &       \\
\arrayrulecolor{black}\bottomrule
\end{tabular}
\end{table}

$S_1$ has one input channel and 96 output channels. $S_1$ begins by splitting the input audio signal into 96 identical copies which then feed into the 96 channels.  These 96 channels are processed in parallel and independent of each other. Each channel starts with FIR (finite impulse response) filtering, followed by the application of gain and bias, then HWR, and finally a low-pass FIR filter and sub-sampling with a factor of four.

$S_2$ through $S_{13}$ have 96 input channels and 96 output channels. The filtering layer of these sections can be described as a full matrix of filtering.  That is, $96^2 = 9216$ filters are used to produce 96 filtered versions of each input channel.
Then each of the 96 output channels is formed by summing one filtered version of each input channel (96 signals in each sum).
The remaining layers of $S_2$ through $S_{13}$ are the same as those in $S_1$, although the sub-sampling may use  a factor of two, three, or four. In each of these layers the 96 channels are processed in parallel and independent of each other. $S_6$ and $S_9$ start with zero padding (appending one zero at the end of the signal) to allow sub-sampling by a factor of two at the end of the section.

\subsection{Per-Function Operation}
A time-domain description of the operations says that WAWEnets replace samples with weighted local averages (convolutions), add up signals from 96 different channels, scale and shift all samples in a channel uniformly (batch normalization), replace negative samples with zero (HWR), and replace blocks of samples with their average value (average pooling). A frequency-domain description provides more insight, so we next describe how these functions change the spectra of the signals as they move through the network. We describe separately changes to the DC component of the signal spectra and changes to the other components. This distinction is key to the description of the overall operation that follows.

The second-order FIR filters have three unconstrained real coefficients, resulting in either a pair of complex-conjugate zeros, or two zeros on the real line. The result is gentle spectral peaks and gentle or deep spectral nulls, depending on the location of the zeros relative to the unit circle.
Forty percent of these filters are lowpass, 29\% highpass, 19\% bandstop, and 12\% bandpass. These proportions change by less than 1\% between \wawenetmultiobj, \wawenetmultisubj, and \wawenetsubj. This commonality is consistent with the fact that both \wawenetmultisubj\ and \wawenetsubj\ use the weights from
\wawenetmultiobj\ as their initial state during the training process.

\begin{figure}
   \centering
   \includegraphics[width=85mm]{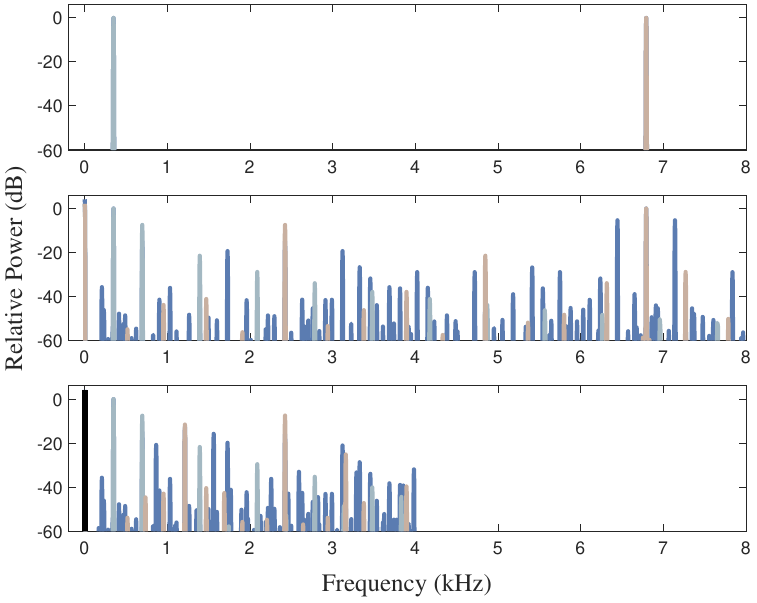}
  \caption{Top panel shows spectra of two tones (345 Hz, shown in light blue, and 6789 Hz, shown in gold), middle panel shows the result of half-wave rectification with zero bias, bottom panel shows the additional effect of average pooling (pooling factor is 2). Colors emphasize non-linearity: light blue shows spectral components that are produced by the 345 Hz tone alone, gold shows components produced by the 6789 Hz tone alone, blue shows components that only appear when both tones are present. A significant DC component will be produced in any of these three cases, and it is shown in black.}
   \label{fig:spectra}
\end{figure}

The gain function simply scales all values of the spectrum (DC and non-DC) by a single value. The bias function adjusts only the DC value of the spectrum and leaves the rest unchanged. The effect of HWR is controlled by the bias. If the bias forces all samples to be negative, HWR removes the signal.  If the bias forces all samples to be positive, HWR does nothing.  If the bias is such that the signal is bipolar, the most common and prominent spectral effect is the creation of new spectral components, thus increasing the spectral density. 
An example is given in the middle panel of Fig. \ref{fig:spectra}.  
The effect of HWR on the DC value of the signal depends strongly on the signal.

The FIR filtering that precedes the subsampling is length $m$ with all coefficients equal to $m^{-1}$ ($m =$ 2, 3, or 4). These are low-pass filters and each one has a perfect null at any frequency that would alias to DC so the subsequent subsampling cannot change the DC value of the spectrum.
(This is consistent with the fact that averaging cannot change the DC value of a signal.)  The subsampling will remove all spectral content above the new Nyquist frequency and can produce aliasing at any non-DC frequency below the new Nyquist frequency. The aliasing is significant because these FIR low-pass filters are very short and have responses that are far from the near brick-wall responses needed to achieve alias-free sub-sampling. For example, when $m=$2 spectral components just above the new Nyquist frequency are aliased to those just below the new Nyquist frequency with only 3 dB of attenuation.
Aliasing involves addition of complex values, so aliasing may reduce or increase the original spectral magnitudes depending on the relative phases of the two addends. The bottom panel of Fig. \ref{fig:spectra} shows an example.

Note that in conventional sample rate reduction, a filter calculates one output sample for each input sample, then subsampling retains every $m^{th}$ sample.  The avgpool function integrates filtering and subsampling so that only every $m^{th}$ sample is calculated.  Because the filtering is FIR and the filter length matches the downsampling factor, this unconventional approach produces the same results as the conventional approach.

\subsection{Overall Operation}
We can view the end-to-end mapping from audio signals to the 96-D latent space as 96 individual (but coupled) signal processors. The job of these processors is to shorten the signals and to strategically shape and move relevant spectral information to DC. In $S_{13}$ the length-3 filtering and 3-to-1 sub-sampling (the avgpool layer) serve to extract the DC value of a signal that has 3 samples. The ensemble of the 96 DC values defines a vector in the latent space and this vector is then mapped to a final output value by $S_{14}$.

\begin{figure}[t!] 
   \centering
   \includegraphics[width=65mm]{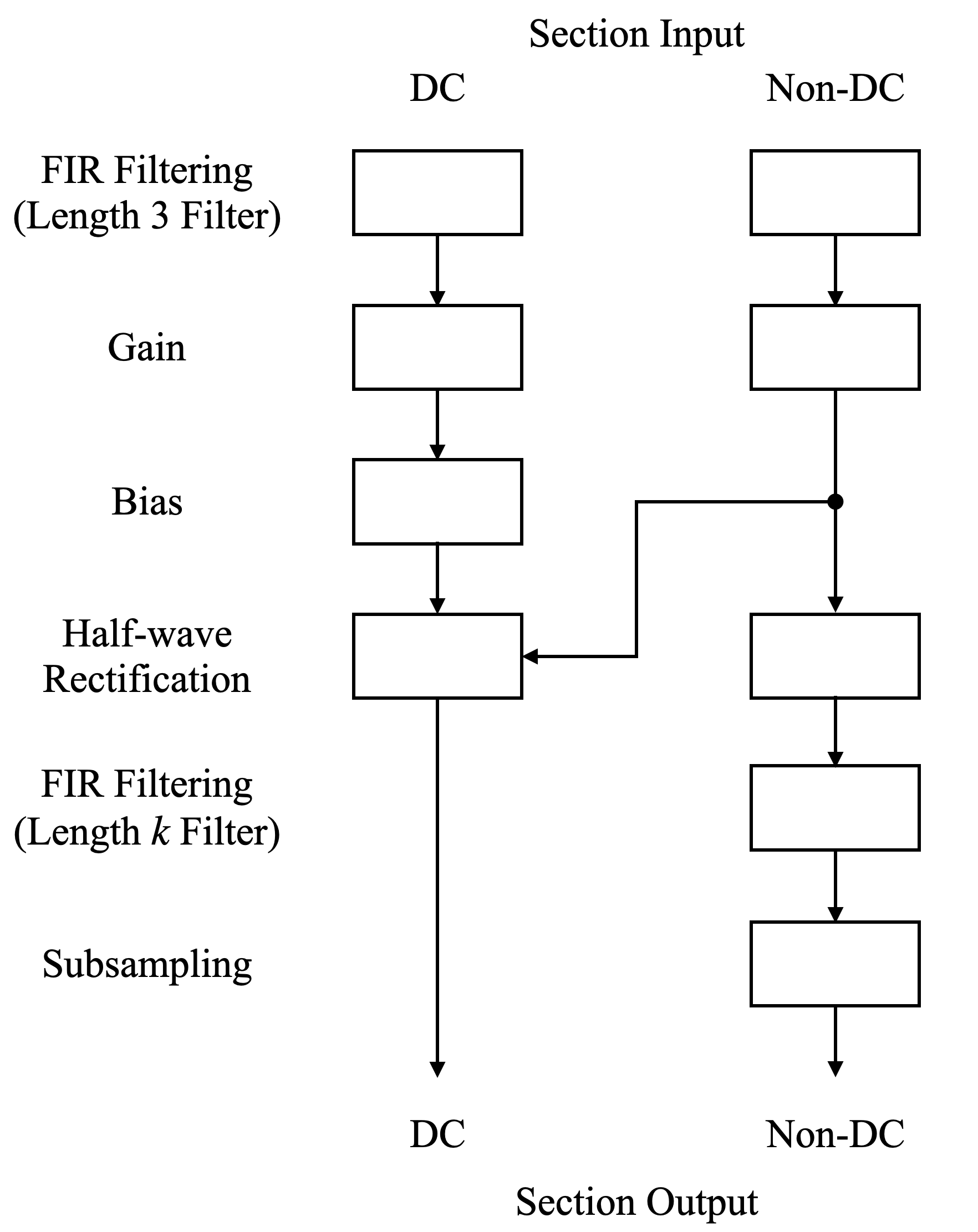}
   \caption{In each section, four of the six functions change DC component and five of the six change non-DC components. Non-DC components affect the DC component \emph{only} in the half-wave rectification function. Sections sequentially move non-DC information to DC.}
   \label{fig:flow}
 \end{figure}

The particulars of shaping spectral information and moving it to DC are shown graphically for one section in Fig. \ref{fig:flow}. In each section, non-DC spectral values of the signals are modified by FIR filtering, gain, HWR, the pooling FIR filter, and sub-sampling. The DC spectral values of the signals are modified by just four of the six functions. In the FIR filtering, gain, and bias functions, the modification of the DC value is determined solely by the processor.  These modifications are independent of the signal itself. But in the HWR function the modification of the DC spectral component is driven by the non-DC spectral components. The HWR is the stage where relevant non-DC spectral information is strategically moved to DC.

For example, consider a speech signal and noise signal each with a DC value of zero.  When added, the noisy speech signal still has a DC value of zero. But after HWR, the original and noisy speech signal can have very different DC values, and these DC values can thus serve to indicate that noise was present in the speech signal. This is a very simple example, and by using many sections of intricate spectral shaping and folding, this processor can also accurately assess a broad array of much more nuanced perturbations to speech signals. 

Fig. \ref{fig:flow} emphasizes the fact that every function modifies either the DC component or the non-DC component, or both, and that the HWR is the sole function where non-DC information influences the DC value. The input audio signal will typically have a DC value near zero, but, as the processing continues, the spectral shaping and the flow of information to DC at HWRs results in DC values that describe important characteristics of the original audio signal.

\begin{figure}[t!] 
   \centering
   \includegraphics[width=85mm]{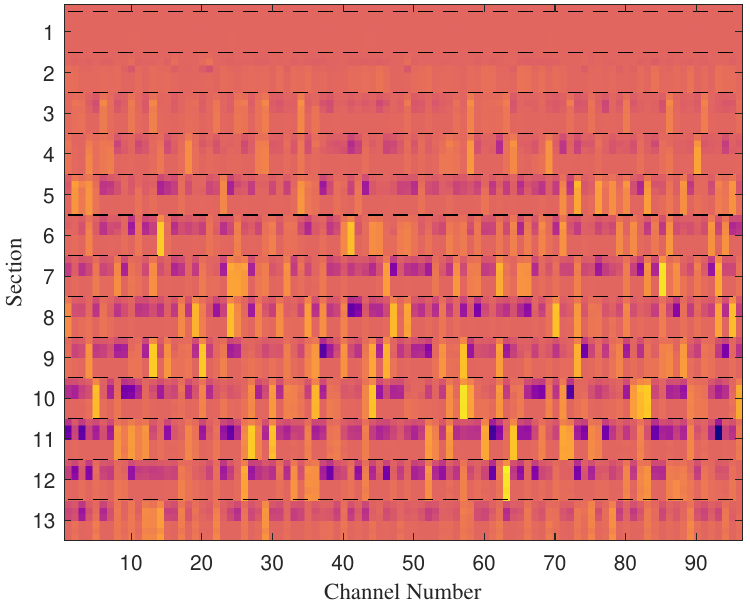}
   \caption{Example DC values (dark purple for large negative, bright yellow for large positive) in the first 13 sections (shown as labeled macro-rows) and all 96 channels (shown as columns) of \wawenetsubj.The six rows within each section show the DC output value of six signal processing functions (FIR, gain, bias, HWR, FIR, sub-sampling, in that order) in that section. The DC value of the input signal is replicated 96 times above $S_1$. }
   \label{fig:dcflow}
 \end{figure}

Fig. \ref{fig:dcflow} gives a visual example of DC values in the first 13 sections of \wawenetsubj\ (1 input signal plus the outputs of 6 SP functions $\times$ 13 sections gives 79 rows) for each of the 96 channels. Input speech is shown at the top of the figure and DC values there are all zero, as expected. As signals move through the network (down the figure) DC values build in various channels for part or all of a section. No continuous downward ``flow'' of DC appears because channels are fully connected with each other in every section and channel numbers are arbitrary. The most dramatic distribution of DC values is in the second half of the network (lower half of the figure), and this distribution then moderates to form the output.

\begin{figure}[h] 
   \centering
   \includegraphics[width=85mm]{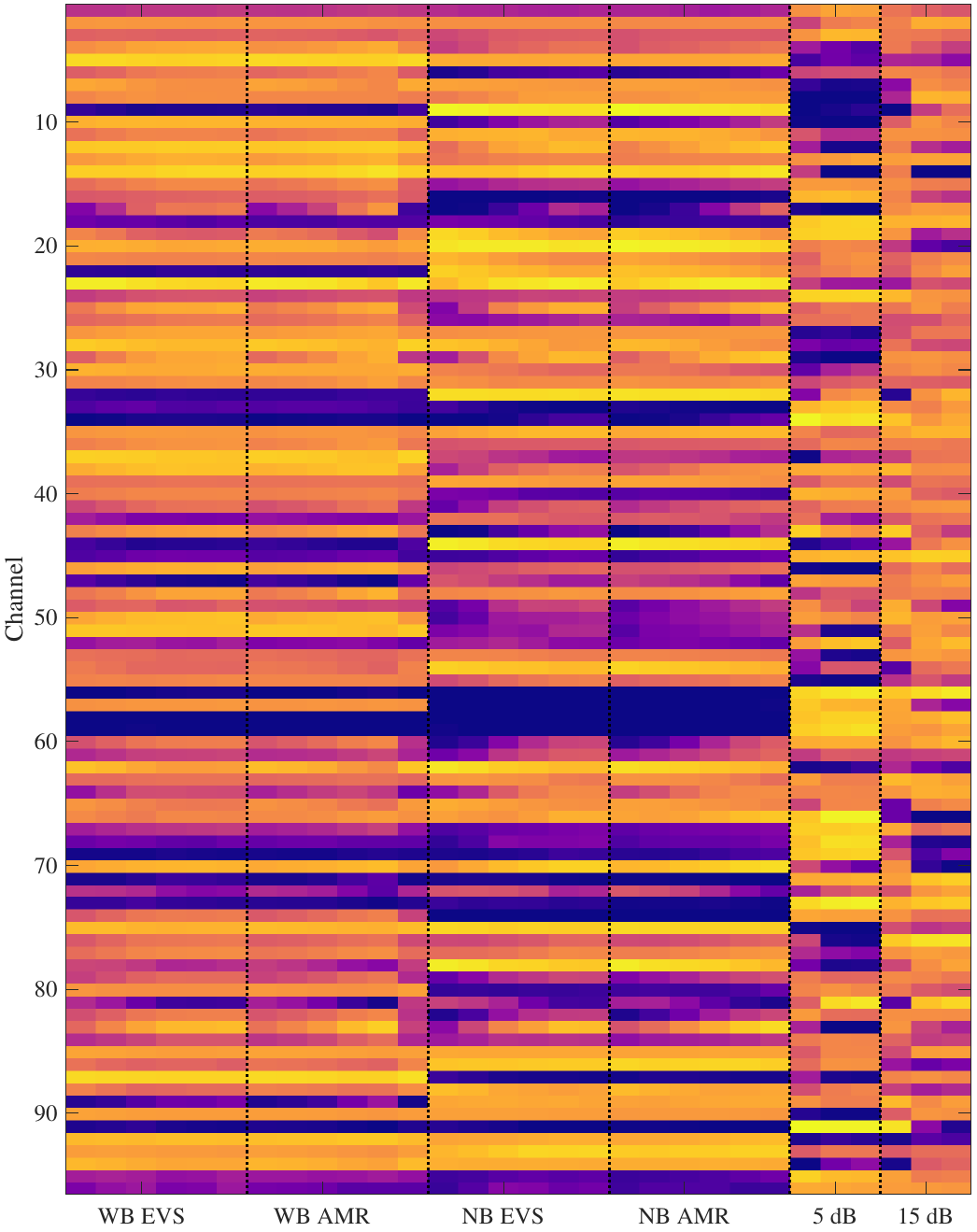}
   \caption{96 outputs of $S_{13}$ for 30 conditions. From left to right, six increasing bit-rates for each of four codecs, then three noise suppression thresholds for each of two SNRs. Example observations: channel 32 responds strongly to WB vs NB and also to noise suppression thresholds; channel 17 responds to codec bit-rates and to noise levels.}
   \label{fig:conds}
 \end{figure}

In effect, the wideband input speech signal passes through 96 parallel coupled signal processors (composed of $S_1$--$S_{13}$) and the 96 DC values of the 96 output signals form a vector that is then mapped by $S_{14}$ to an estimate of some quality of that speech signal.
Fig. \ref{fig:conds} shows examples of these 96-D \wawenetmultiobj\ outputs for 30 different conditions (1,320 segments averaged for each condition). These examples show how different dimensions respond to different attributes while they work together to produce an estimate of speech quality or intelligibility. 

\section{Conclusion}
WAWEnets are no-reference wideband audio waveform evaluation networks that process waveforms directly into evaluations of those waveforms.
We have trained and evaluated multiple WAWEnets, and this work is based on 334 hours of speech in 13 different languages produced by more than 1000 different talkers.
We have demonstrated that \wawenetmultiobj\ can produce speech quality and intelligibility estimates that agree (correlations of 0.92 to 0.96) with values from seven established FR objective estimators.
\wawenetmultisubj\ demonstrates that this architecture is useful for estimating both objective and subjective speech qualities at the same time.
\wawenetsubj\ agrees with the subjective MOS values of the IUB dataset with per-segment correlation  0.973, improving upon top results from other current approaches.
These three sets of results show that WAWEnets can provide useful estimations of audio qualities when trained on either large or small datasets that contain either real or simulated speech impairments.

The number of parameters in \wawenetmultiobj~is only 0.4\% of the number in wav2vec and 4.5\% of the number in TorchAudio-Squim.  And the operations required to run \wawenetmultiobj~are 3.3\% of those required for wav2vec and 2.7\% of those required for TorchAudio-Squim.  This makes \wawenetmultiobj~quite lightweight and energy-efficient relative to those two alternatives.  In these respects, \wawenetmultiobj~ is comparable to NISQA-DIM.  In addition, our experiments found \wawenetmultiobj~and NISQA-DIM had matched correlations, and \wawenetmultiobj~ had slightly lower RMSE than NISQA-DIM. This demonstrates that allowing CNNs to learn to extract the needed information directly from waveforms (as in WAWEnets) is a perfectly viable alternative to using hand-crafted features as network inputs (as in NISQA-DIM).
We expect that \wawenetmultiobj~will generate accurate predictions of speech qualities for applications where speech impairments resemble those represented in our wide-ranging training dataset.
In addition, we expect they would be useful for narrower applications (for example, measuring reverb present in a room) if trained to do so.

WAWEnets' small size make them approachable for applications where computing power is at a premium.
This fact, combined with the fact that they work directly on waveforms, means WAWEnets are uniquely suited for endpoint monitoring.
This monitoring could be logged along with other relevant information to support retrospective diagnostics, address questions of compliance with service level agreements, or enable studies of which configurations or conditions create changes that would be audible to users at the endpoint and which would be transparent. Results could also be passed back upstream to inform network elements of user experience so they could adapt as needed.  And WAWEnets can also be used offline to efficiently analyze large batches of recorded speech files.

WAWEnets are distinct from the vast majority of alternatives because they apply ML directly to speech waveforms instead of applying it to features extracted from speech waveforms. This gives WAWEnets access to all the information in the waveforms and allows them to, in effect, generate the best features for the task internally, rather than having potentially sub-optimal features mandated externally. Our work shows that this is indeed a viable approach.

At present, WAWEnets operate only on 3-second segments of speech sampled at 16,000 samples/second. We have proposed (see Section \ref{sec:disc}) several strategies to extend WAWEnets to wider bandwidths and longer signals as labeled data becomes available.

WAWEnets are composed of six common signal processing functions and, as expected, the non-linear functions (bias and HWR) are critical.  These functions move spectral information (shaped by FIR filtering and gains) to DC so that the DC values of the final, very short signals provide a 96-D description of the original, much longer signals. 

We encourage further experimentation and development with WAWEnets, especially in the case that impairments in your dataset differ significantly from impairments in our dataset. To that end, please visit the software repository at https://github.com/NTIA/WEnets where we provide the weights derived in this work and code suitable for training WAWEnets on your dataset.

\section*{Acknowledgments}

Sincere thanks to Gabriel Mittag, Sebastian M{\"o}ller, and the members of the Quality and Usability Lab of the Technische Universit\"at Berlin for creating and sharing the TUB dataset; to Zhuohuang Zhang, Donald Williamson, and the members of the Audio, Speech, and Information Retrieval Group at Indiana University Bloomington for creating and providing the IUB dataset; and to Michael Chinen, Andrew Hines, and Jan Skoglund for support with ViSQOL 3.0.

\ifCLASSOPTIONcaptionsoff
  \newpage
\fi

\bibliographystyle{IEEEtran}

\bibliography{sources_cleaned}

\begin{biography}{Andrew A. Catellier}
(Senior Member, IEEE) attended the University of Wyoming, where he earned a B.S. in Electrical Engineering in 2006 and an M.S. in Electrical Engineering in 2007. 
Between 2007 and mid 2023, Andrew conducted traditional and crowd-sourced subjective experiments for audio and video signals, and explored objective speech quality assessment using ML techniques at the Institute for Telecommunication Sciences (ITS) in Boulder, Colorado. 
From 2017 until 2022, he worked to facilitate crop yield measurements, forecasts, and classification via remote sensing, computer vision, and ML at GeoVisual Analytics from 2017 until 2022. 
Starting in 2022, Andrew leveraged computer vision and ML to validate identification documents and identity at Nametag, Inc.
\end{biography}

\begin{IEEEbiography}{Stephen D. Voran}
(Senior Member, IEEE) received the B.A. degree in Mathematics from Carleton College, Northfield, MN, in 1985, and the M.S. degree in Electrical Engineering
from the University of Colorado, Boulder, in 1989. Since 1990 he has been with the Institute for Telecommunication Sciences, Boulder, Colorado and has been contributing to signal-processing–based advances in objective assessment of telecommunications speech quality and intelligibility, as well as subjective audio testing, speech coding, and audio quality enhancement.

\end{IEEEbiography}

\EOD

\end{document}